%% file: main.tex
\documentclass[sigconf, nonacm]{acmart}

\AtBeginDocument{%
  \providecommand\BibTeX{{%
    \normalfont B\kern-0.5em{\scshape i\kern-0.25em b}\kern-0.8em\TeX}}}

\setcopyright{acmcopyright}
\copyrightyear{2022}
\acmYear{2022}
\acmDOI{10.1145/xxxxxxx.xxxxxxx}

\acmConference[WiSec '22]{WiSec '22: 15th ACM Conference on Security and Privacy in Wireless and Mobile Networks}{May 16--May 19, 2022}{San Antonio, Texas, USA}
\acmBooktitle{WiSec '22: 15th ACM Conference on Security and Privacy in Wireless and Mobile Networks, May 16--May 19, 2022, San Antonio, Texas, USA}
\acmPrice{15.00}
\acmISBN{978-1-4503-XXXX-X/21/06}

\usepackage{tabularx}
\usepackage{subcaption}
\usepackage[ruled]{algorithm2e}
\usepackage{dsfont}
\usepackage{makecell}
\usepackage{cleveref}
\usepackage{siunitx}
\ifdefined\unit\else
  \ifdefined\NewCommandCopy
    \NewCommandCopy\unit\si
  \else
    \NewDocumentCommand\unit{O{}m}{\si[#1]{#2}}
  \fi
\fi
\usepackage{multirow}
\usepackage{amsmath}
\usepackage{amsfonts}
\usepackage{pdfrender}
\usepackage{tablefootnote}
\usepackage{natbib}
\usepackage{enumitem}
\usepackage{balance}
\usepackage{changepage}
\usepackage{eucal}
\usepackage{bm}

\newcommand{\bfpara}[1]{\vskip 1ex \noindent \textbf{#1.}}
\newcommand{\ignore}[1]{{}}

\newenvironment{myfont}{\fontfamily{pbk}\selectfont}{\par}
\DeclareTextFontCommand{\textmyfont}{\myfont}

\newcommand{\code}[1]{\texttt{#1}}
\definecolor{codegray}{gray}{0.9}

\begin{document}


\title{Towards an AI-Driven Universal Anti-Jamming Solution with Convolutional Interference Cancellation Network}

\author{Hai N. Nguyen}
\email{nguyen.hai@northeastern.edu}
\affiliation{%
  \institution{College of Computer Sciences, Northeastern University}
   \city{Boston}
   \state{Massachusetts}
   \country{USA}
}

\author{Guevara Noubir}
\email{g.noubir@northeastern.edu}
\affiliation{%
  \institution{College of Computer Sciences, Northeastern University}
   \city{Boston}
   \state{Massachusetts}
   \country{USA}
}


\input{abstract}

\maketitle

\input{intro}

\input{overview}

\input{cnn}

\input{cancellation}

\input{evaluation}

\input{related}


\bibliographystyle{ACM-Reference-Format}
\bibliography{main}

\end{document}

%% file: abstract.tex
\begin{abstract}
Wireless links are increasingly used to deliver critical services, while intentional interference (jamming) remains a very serious threat to such services. In this paper, we are concerned with the design and evaluation of a \textit{universal} anti-jamming building block, that is agnostic to the specifics of the communication link and can therefore be combined with existing technologies. We believe that such a block should not require explicit probes, sounding, training sequences, channel estimation, or even the cooperation of the transmitter. To meet these requirements, we propose an approach that relies on advances in Machine Learning,  and the promises of neural accelerators and software defined radios. We identify and address multiple challenges, resulting in a convolutional neural network architecture and models for a multi-antenna system to infer the existence of interference, the number of interfering emissions and their respective phases. This information is continuously fed into an algorithm that cancels the interfering signal.  We develop a two-antenna prototype system and evaluate our jamming cancellation approach in various environment settings and modulation schemes using Software Defined Radio platforms. We demonstrate that the receiving node equipped with our approach can detect a jammer with over 99\% of accuracy and achieve \ignore{high jamming resilience when operating at }a Bit Error Rate (BER) as low as $10^{-6}$ even when the jammer power is nearly two orders of magnitude (18 dB) higher than the legitimate signal, and without requiring modifications to the link modulation. In non-adversarial settings, our approach can have other advantages such as detecting and mitigating collisions. 
\end{abstract}

%% file: intro.tex
\section{Introduction}

The mobile revolution is today a reality beyond its pioneers' dreams. This is, to a significant extent, due to the progress achieved by wireless communications systems, from increased throughput, to incredible reductions in size and power consumption. The massive integration of wireless in everyday devices, is not only a convenience, it profoundly changed how a variety of systems are designed and operated. Beyond the deluge of IoT devices, connected through wireless links (e.g., smart-homes and wearables), we are experiencing a disappearance of wired links. A plethora of older critical Cyber-Physical System are transitioning to wireless connectivity such as the use of Wireless Remote Terminal Unit (RTU) in the monitoring and control of SCADA systems including the electricity grid and industrial processes. Even aircraft and aerospace systems cannot resist the benefits of wireless connectivity~\cite{Zaghari2020HighTemperatureSS}. Furthermore, emerging CPS technologies such as self-driving cars significantly rely on a multitude of wireless links for their operations. In the last decade, wireless connectivity also underlied notable events for society around the world, from the Arab Spring, to supporting police accountability. 


Jamming remains one of the most serious threats to wireless communications. Wireless links typically require a Signal to Noise Ratio (SNR) in the order of 20 dB to meet the throughput requirements of most systems. On the other hand, jammers do not require sophisticated RF Front-Ends design and basic jamming hardware against Wi-Fi, cellular networks, and GPS devices is a commodity that can be found on the Internet for a few dozens of dollars. Due to a series of incidents, the FCC routinely releases customer advisories cautioning against the import and use of jamming devices~\cite{FCC}, stepped up its education and enforcement effort, rolled out a jammer tip line (1-855-55NOJAM), and issued several fines~\cite{FCC-fine-2022}. Despite the regulation, preventing jamming remains difficult to enforce. Furthermore, wireless softwarization is making jamming potentially a ubiquitous threat, as demonstrated by the nexmon framework~\cite{SchulzWH2016}, and Google Project Zero exploitation of a Wi-Fi chipset firmware~\cite{google-projectzero-wifi}, and recognized in numerous work~\cite{WilhelmMSL11}. Finally, the impact of jamming goes beyond denial of the increasingly critical services relying on wireless communication, it can also be the prelude to more sophisticated attacks such as rogue infrastructure (Wi-Fi and Cellular) and hijacking of physical assets (GPS).

In this paper, we are concerned with the design and evaluation of a \textit{universal} anti-jamming building block, that is agnostic to the specifics of the communication link and can therefore be combined with existing technologies. We believe that such a block should not require explicit probes, sounding, training sequences, channel estimation, or even the cooperation of the transmitter. To meet these requirements, we propose an approach that relies on advances in Machine Learning and the promises of neural accelerators and software defined radios. In a nutshell, we develop a deep learning architecture and models that drive a multi-antenna system to infer the existence of interference, the number of interfering emissions and their respective phases. This information is continuously fed into an algorithm that cancels the interfering signal. We note that wireless communications aim at very low Bit Error Rate (BER typically below $10^{-4}$) while ML systems are typically content with much lower performance. Therefore, the ML estimates require careful processing to avoid abrupt changing in the receiver RF chain. We address these challenges, prototype our techniques, and evaluate them in a variety of scenarios with multiple modulation schemes, demonstrating good performance for a receiver exposed to a jammer that is two orders of magnitude higher than the legitimate signal. For instance, we demonstrate that a receiver integrating our approach can detect the jammer with over 99\% of accuracy and achieve high jamming-resistance with a Bit Error Rate (BER) as low as $10^{-6}$ while the jammer power is 18dB higher than the legitimate signal. We note that our approach does not require changes to the link modulation schemes or reliance on approaches such as spread spectrum. Our contributions can be summarized as follows:

\begin{itemize}
    \item A novel RFML approach based on Convolutional Neural Networks (CNNs) for \textit{universal} anti-jamming and cancelling interference.
    \item A neural network architecture, model, and supporting algorithms for high-accuracy multi-antenna jamming cancellation.
    \item A two-antenna SDR-based prototype receiver that leverages the proposed CNN model and algorithms for jamming cancellation and operates in a modulation-agnostic way (BPSK, QPSK, 8-PSK, 16-QAM).
    \item The system is evaluated in various settings demonstrating the jammer detection accuracy of over $99\%$ and achieving a BER as low as $10^{-6}$ even against jammers that are 18dB more powerful than the legitimate sender. Moreover, our system can achieve over $30$dB gain when operating at a BER under $10^{-4}$ compared to a receiver without our jammer cancellation.  
\end{itemize}

%% file: overview.tex
\section{Problem Statement and Approach}

We first formulate the problem of interest, the communications and adversarial models, and present the underlying theoretical foundations to our jammer cancellation approach. 

\begin{figure}[t]
    \centering
    \subcaptionbox{\label{fig:setup_1}Single-antenna receiver.}{
        \includegraphics[width=.9\linewidth, height=3cm]{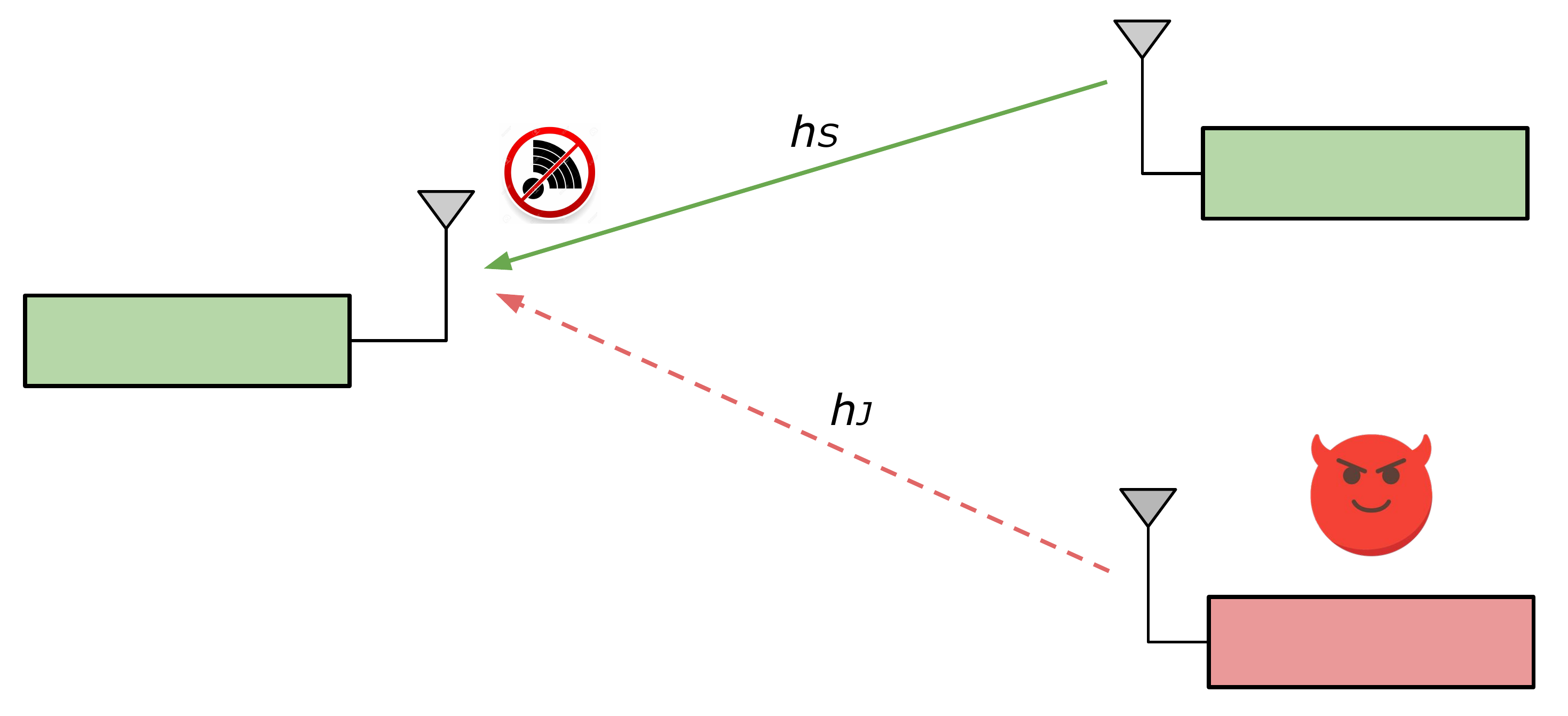}
    }
    \subcaptionbox{\label{fig:setup_2}Two-antenna receiver.}{
        \includegraphics[width=.9\linewidth, height=3cm]{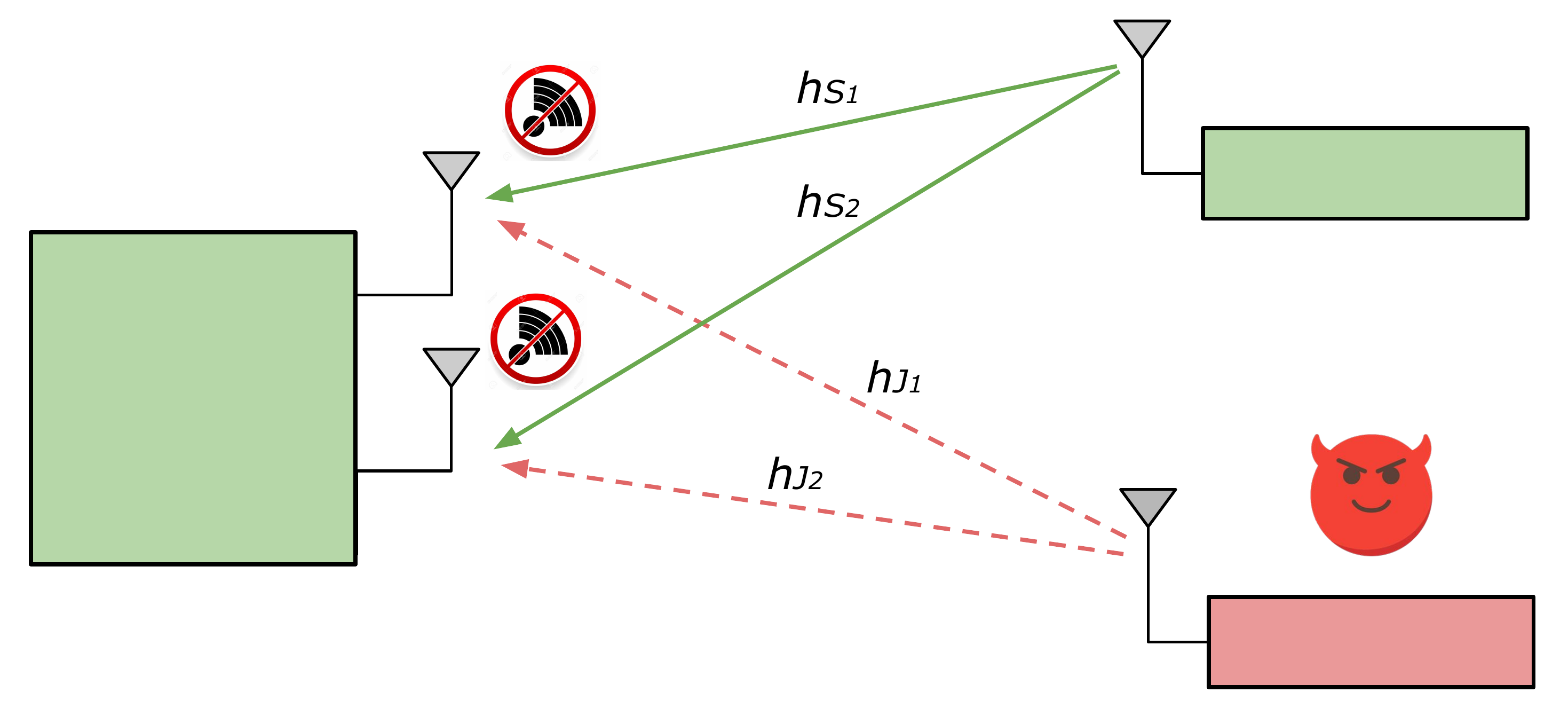}
    }
    \caption{Illustration of a jammer interfering with the communications between two nodes.}
    \label{fig:setup}
\end{figure}

\subsection{Models}
\label{sec:models}

We consider a setup with two legitimate nodes communicating in the presence of an adversary intentionally interfering with the communications. In the following, we define the models for the legitimate nodes, adversary, and channel.

\bfpara{Sender and Receiver Model} For the purpose of this work, we assume that the legitimate nodes are communicating over a pre-agreed channel and link parameters, including the center frequency, bandwidth, modulation scheme. This is a standard assumption and further resilience to jamming can be achieved by allowing the communicating nodes to randomize such parameters (e.g., frequency hopping). We assume that the sender uses a single transmitting antenna, and that the receiver uses two antennas with the same gain to receive signals. We assume that the nodes are \textit{neither} aware of each other's location, nor the location of the jammer. We consider a slow-fading channel, therefore a low-mobility for the involved parties. This concretely means that the channel characteristics do not change abruptly within a packet (e.g., 1 ms). 

\bfpara{Adversary Model} We consider an attacker (jammer) using a single antenna transmitting signals on the same channel as the users, interfering with the legitimate communications (\Cref{fig:setup}). We allows a powerful adversary that already knows the link parameters such as center frequency, bandwidth and potentially other settings.  The jammer is allowed to transmit either random interfering samples, or modulated packets, with a continuous or intermittent pattern (it can also be a reactive jammer). We also assume a similar low-mobility pattern for the jammer so that the channel characteristics do not change abruptly within a packet.

\bfpara{Communication Channel Model} The communicating nodes can use arbitrary modulation and coding schemes and are exposed to the typical additive Gaussian noise (AWGN) in addition to the jammer interference. We assume the channel gains to be fairly stable throughout the considered bandwidth. In our evaluations, we consider differential BPSK, QPSK, 8-PSK, and 16-QAM modulations for the communicating nodes. We evaluate our system in both over-the-air setups, and using RF coax cable. The cable setup is for reproducibility and in order to systematically and extensively assess the performance of the approach over a range of three orders of magnitude of powers (35~dB), and multiple phase offsets.  

\subsection{Theoretical Foundations  and Approach} 
\label{sec:approach}

\bfpara{Jamming Fundamentals} Consider a sender and a receiver communicating over a slow fading channel. It is standard to model a conventional system using a single antenna, as receiving a signal $R$ (represented as a complex number) that comprises the transmitting signal $S$, adjusted to account for the channel gain $h$, and additive white Gaussian noise $N$:

\begin{equation}\label{eq:recv_sig}
    R=h S + N
\end{equation}

In the absence of interference, the quality/capacity of such link is determined by the Signal-to-Noise Ratio (SNR), which is proportional to $\frac{\left|h\right|^2}{\left|N\right|^2}$ (where $|.|$ denotes the complex norm). Assuming that additive noise is constant over time, the SNR only depends on $\left|h\right|$. The receiver achieves better Bit Error Rate (BER) when the channel gain $\left|h\right|$ is higher, which is reflected in a higher SNR.

Now assume the presence of an adversary (jammer), who knows the frequency channel that the legitimate nodes are operating over. The adversary transmits a "jamming" signal $J$ that interferes with the legitimate signal $S$ (as shown in \Cref{fig:setup_1}). The receiving signal now becomes:
\begin{equation}\label{eq:recv_sig_jam}
    R=h_{S} S + h_{J} J + N
\end{equation}
where $h_{S}$ and $h_{J}$ are the channel gains corresponding to the sender and the jammer, respectively. The decodability of the legitimate signal is now dependent on the Signal-to-Interference-and-Noise Ratio (SINR) which is proportional to $\frac{\left|h_{S}\right|^2}{\left|h_{J}J\right|^2+\left|N\right|^2}$. When the jamming power is considerably high relatively to the channel noise, $\left|h_{J}J\right| \gg \left|N\right|$,  the SINR can be approximated as proportional to the channel gain ratio $\frac{\left|h_{S}\right|^2}{\left|h_{J}J\right|^2}$. As the interference becomes stronger, $\frac{\left|h_{S}\right|}{\left|h_{J}J\right|}$ is subsequently smaller and the legitimate signal $S$ becomes undecodable.

\bfpara{Approach} To remove the jamming component from the received signal $R$, in a single-antenna, is challenging without having  control over the jammer, knowing the jamming signal, or resorting to other dimensions to evade the jammer (e.g., as in spread spectrum).  Our approach instead relies on two antennas at the receiving node, each collects a copy of the transmitted signals (subject to different channel gains):
\begin{align}
\begin{split}
    R_{1} = h_{S_1} S + h_{J_1} J + N_1 \\
    R_{2} = h_{S_2} S + h_{J_2} J + N_2 \label{eq:recv_sig_jam_multi}
\end{split}
\end{align}
Considering a jammer significantly above the noise, the cancellation is achieved using the formula:
\begin{equation}\label{eq:jam_cancel}
    R_1 - p_1 R_2 = p_2 S 
\end{equation}
where $p_1=\frac{h_{J_1}}{h_{J_2}}$, and $p_2=h_{S_1}-p_1 h_{S_2}$. If the new channel gain $p_2$ of signal $S$ is sufficiently large, we can decode $S$ and achieve a good BER.

The main challenge is how to estimate parameter $p_1$ correctly. Here, we emphasize that traditional MIMO approaches rely on training sequences and sounding procedures (cooperatively between the transmitter and receiver) to estimate the channel gain. Such approach is not applicable in this case, because the receiver cannot control the jamming signal. We first reformulate $p_1$ in the polar representation $\frac{\left|h_{J_1}\right|}{\left|h_{J_2}\right|}e^{j(\phi_{J_1}-\phi{J_2})}$. To find $p_1$, we are required to estimate the \textit{amplitude ratio} $A_J =\frac{\left|h_{J_1}\right|}{\left|h_{J_2}\right|}$ and the \textit{phase shift} $\Delta_{\phi_J}$:
\begin{equation}\label{eq:phase_shift}
 \Delta_{\phi_J} = \phi_{J_1}-\phi_{J_2}
\end{equation}
We use two different approaches to estimate $A_J$ and $\Delta_{\phi_J}$. For the \textit{amplitude ratio}, we rely on the fact that the parameter is proportional to the square root of the ratio of jamming power received at the antennas. On that account, we estimate using the measured power in the periods before and during the collision (~\Cref{sec:amp_est}).  To estimate the \textit{phase shift}, we developed a lightweight, yet powerful Convolutional Neural Network (CNN) that directly estimates  from the $I/Q$ RF samples. The ability of CNN to analyze and infer diverse complex data has been investigated and utilized in various areas~\cite{oshea16, VGG-2014,dcnn_language, cnn_speech_recog}. Estimating the phase shift involves extracting and synthesizing low-level patterns of the original jamming signal embedded in the RF samples. In conventional wireless systems design, such patterns are extracted through signal processing filters, which makes the layers of convolutional filters of CNN an ideal candidate for the task. Our CNN not only can disentangle the collision and estimate the phase shifts for both legitimate and jamming signals, but also can infer if the estimations correspond to data-containing signals or just noise. The latter allows us to detect the presence of a jammer and to distinguish between the three possible states of the communication channel: (1) When the channel is clear, (2) when only the jammer or the sender is transmitting, and (3) when the communicating nodes are being interfered with by the jammer. To the best of our knowledge, our work is the first that considers CNN as a multi-functional approach for detecting and cancelling jammers. More details of the approach are presented in~\Cref{sec:cnn} and~\Cref{sec:cancel}.

However, we note that multi-antenna jammer nulling has intrinsic limitations. Even with accurate estimates of the amplitude ratio and phase shift, it is not guaranteed that jamming cancellation can fully recover the original signal. As we mentioned earlier, removing $J$, results in signal $S$ being subject to an update gain value $h_{S_1}-p_1 h_{S_2}$ where $p_1=\frac{h_{J_1}}{h_{J_2}}$. This gain becomes small when $\frac{h_{S_1}}{h_{S_2}} \approx \frac{h_{J_1}}{h_{J_2}}$, equivalently $\Delta_{\phi_S} \approx \Delta_{\phi_J}$ i.e. the separation between the phase shifts corresponding to channel gains of signals $S$ and $J$ is small:
\begin{equation}\label{eq:phase_separation}
 Sep_{\Delta_\phi}=\left|\Delta_{\phi_S}-\Delta_{\phi_J}\right| \approx 0
\end{equation}
This derives from the intrinsic limitation of multi-antenna system to not be able to distinguish between two emitters that are aligned with the receiver. In the later sections, we will show the impact of parameter $Sep_{\Delta_\phi}$ to the jamming cancellation approach through extensive experiments for Bit Error Rate evaluation.

%% file: cnn.tex
\section{Jamming Detection and Phase Shift Estimation}\label{sec:cnn}
When the communication is interfered by jammer, estimating the phase shift $\Delta_{\phi_J}$ (in~\Cref{eq:phase_shift}) is challenging without explicit information about the jamming signal. Firstly, the received signal introduces an entanglement of legitimate and jamming signals. Secondly, the constructive and destructive effects of multi-path propagation on the signal's phase is typically unpredictable. Our approach to address this challenge centers around a fast and accurate convolutional neural network (CNN) which can estimate the phase shift precisely as well as recognize the current state of the channel (as outlined in~\Cref{sec:approach}) and detect the jammer. We are inspired by CNN's capability of extracting relevant low-level features from various types of data such as visual~\cite{VGG-2014, resnet16}, speech~\cite{cnn_speech_recog}, RF~\cite{oshea16}, and text~\cite{dcnn_language}.\ignore{ Moreover, convolution operation, which is the atomic computational unit of CNN, has been widely used in various tasks of wireless communications such as estimation and filtering, that drives our expectation for using CNN to capture channel characteristics in the variations of amplitude and phase of RF samples.} In this section, we present our design of the CNN, as well as the components for training the neural network specially for the jamming detection and cancellation. 

\subsection{Challenges and Goals of The Design}
As outlined in~\Cref{sec:approach}, our main goal is to accurately estimate the phase shift $\Delta_{\phi_J}$ of the jamming signal $J$. However, to account for all possibilities and ensure the cancellation is done on the right signal (the jamming signal $J$), we first analyze the challenges and define the goals for our CNN design.

\bfpara{Challenges} While developing the convolutional neural network (CNN) model for phase shift estimation, we encountered two main challenges . \textit{\textbf{First,}} with a single estimation, it is hard to guess whether the phase shift associates with the legitimate or the jamming signal. This is especially more confusing when an adversary tries to mimic the legitimate communication, e.g., by using the same modulation. As a result, in the worst case scenario, we can instead inadvertently cancel the legitimate signal. \textit{\textbf{Second}}, the receiver does not know the current state of the communication channel, i.e., how many transmitters are concurrently using it. This could lead to another catastrophic scenario when the sender is transmitting without being interfered, while the receiver still believes that a collision is happening and unintentionally removes the signal using the estimated phase shift. 

\bfpara{Goals} We defined two goals for the design of that target CNN to address the above challenges. \textit{\textbf{First}}, the phase shift estimations for both the legitimate signal $S$ and the jamming signal $J$ are required instead of for the latter only. This is intuitively possible to achieve with a CNN since that signals $S$ and $J$ are typically incoherent, therefore the unique features of both signals can be extracted by stacks of trained convolutional filters. \textit{\textbf{Second}}, for each estimated phase shift, we also infer whether it is the estimation of a data-containing signal or of noise. As such, the neural network outputs a confidence value along each inferred phase. This indicates the presence of a signal if the confidence is larger than 0.5, and indicates noise otherwise. We reckon that this \textit{signal detection} ability can help solve a more general class of problems than an approach of estimating the number of emitters, because the former can identify the phase shift of signals when there is one emitter while the latter cannot. This is useful to track the jamming signal and identify the type of jammer. These capabilities allow us to recognize the current state of the communication channel, to \textbf{\textit{detect the jammer}} (i.e. \textit{when both signal detection outputs indicate a signal}), and to avoid the chance of accidentally removing the legitimate signal without any means to recover.

\subsection{Neural Network Architecture}
\begin{figure}
    \centering
    \includegraphics[width=.9\linewidth, height=9.8cm]{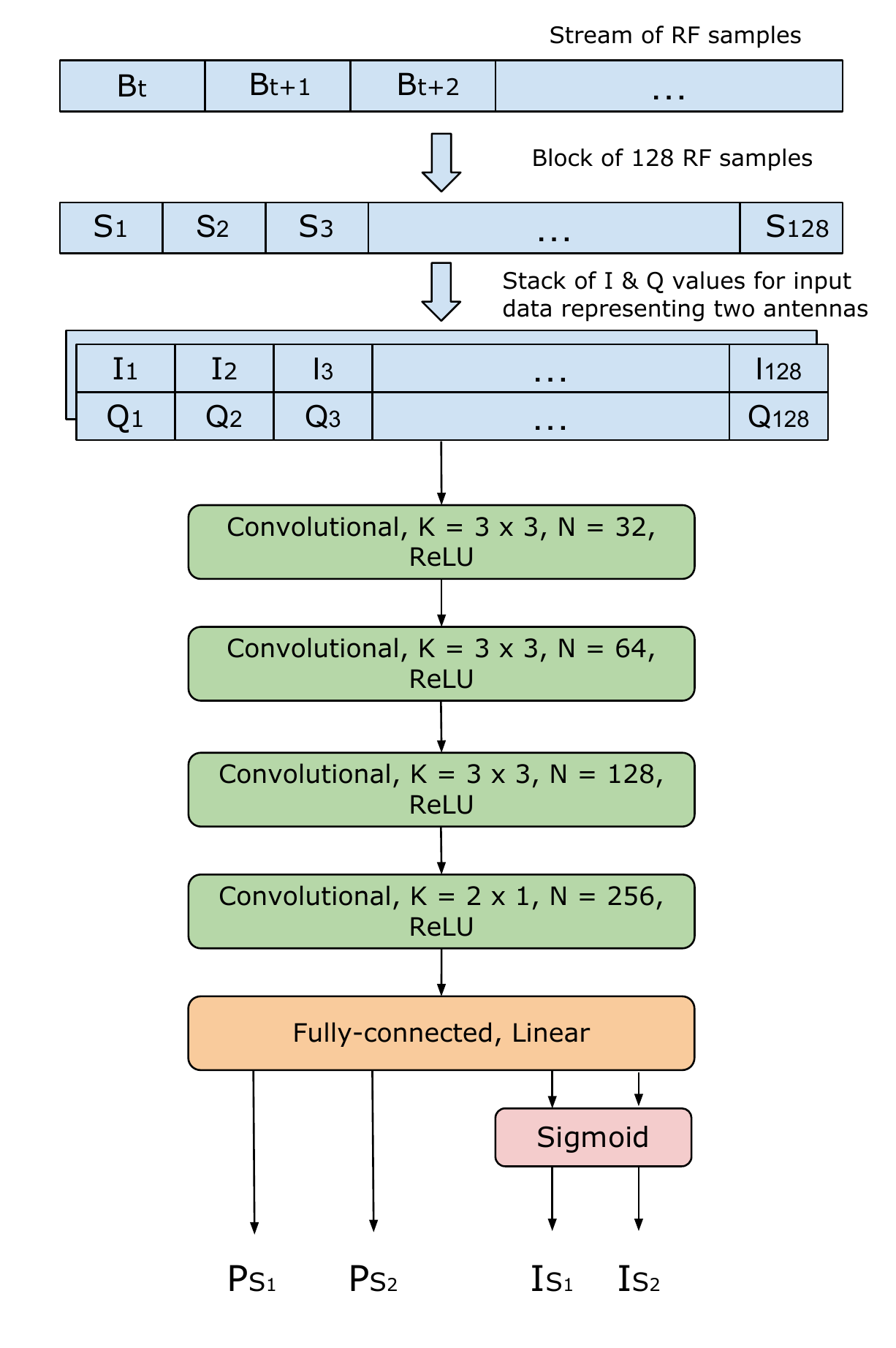}
    \caption{RF data representation and the CNN architecture for phase shift estimation and signal detection. $K$ is the filter size and $N$ is the number of filters in a convolutional layer.}
    \label{fig:cnn}
\end{figure}

Our development of the CNN started with defining the input layer for RF data. Naturally, we want to avoid feeding a very long stream of RF samples to the CNN at once due to the heavy computational cost. To address this, we divide the stream of $I/Q$ samples into blocks of a fixed length $M$  (In our implementation, $M = 128$ samples). Then, we transform each block into a $2 \times M$ matrix where the first row comprises the In-phase (I) values and the second row comprise the Quadrature (Q) values of $M$ RF samples (shown in~\Cref{fig:cnn}). It is noted that this process is done for both antennas at the receiver. Finally, we stack the matrices of the two antennas along the third dimension to form the $2 \times M \times N$ real-valued tensor as the input of our CNN. 

We have considered several possible architecture designs of the CNN and converged on an optimized CNN structure that achieves good performance in terms of processing speed and estimation correctness (See discussion in later sections). The architecture of our CNN is illustrated in \Cref{fig:cnn}, in which a stack of three convolutional layers with kernel size of $3 \times 3$ is followed by a $2 \times 1$ convolutional layer and a fully-connected layer.  $3 \times 3$ convolutional layer is a popular Deep Learning technique and has been used as the building block for state-of-the-art Deep Learning (DL) architecture such as VGG \cite{VGG-2014} and ResNet \cite{resnet16}. As $3 \times 3$ convolution is especially effective for extracting low-level features in different local regions of the data, we expect the neural network to find robust, unique features and patterns of the colliding signals from the variation in the amplitude and phase of contiguous RF samples, and to learn from them effectively. On the other hand, we use the $2 \times 1$ convolutional layer with the sample-wise combining of I \& Q channels to gain the high-level semantics of angular distance for estimating the phase shifts. Rectified Linear Unit (ReLU) activation is used for the convolutional layers because it is computationally efficient and more effective against the vanishing gradient problem~\cite{relu_2011}.

The fully connected layer synthesizes the output from previous layers and makes predictions. The outputs $P_{S_1}$ and $P_{S_2}$ estimate the phase shifts for legitimate/interference signals, while $I_{S_1}$ and $I_{S_2}$ detect whether the corresponding phase shift estimations come from a signal or noise. (Again, 1 implies real signal while 0 implies noise). Sigmoid activation is used for $I_{S_1}$ and $I_{S_2}$ to limit the values in the range $[0, 1]$, while linear activation is used for $P_{S_1}$ and $P_{S_2}$. We emphasize that as $P_{S_1}$ and $P_{S_2}$ cannot be used interchangeably, we distinguish them by having $P_{S_1}$ learn the smaller phase shift while $P_{S_2}$ learns the bigger one.

\subsection{Data Collection}\label{sec:data_collection}
\begin{figure*}
    \centering
    \includegraphics[width=\linewidth]{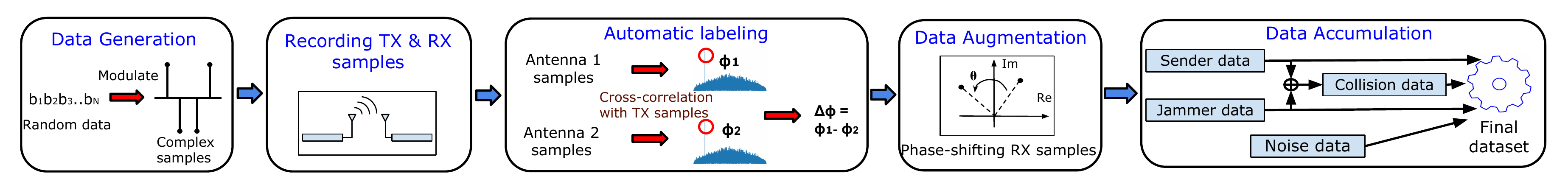}
    \caption{The procedure of building dataset for jamming detection and cancellation.}
    \label{fig:dataset}
\end{figure*}

Having a large and carefully labeled dataset is a critical requirement to train a good neural network model. Unfortunately, data labeling is normally done by manual labor, which requires domain knowledge and takes significant amount of time and efforts. Moreover, open datasets of real RF emissions for jamming cancellation remains absent in the literature. To address this challenge, we devise an efficient multi-stage approach to build a large training dataset for our CNN model, as depicted in \Cref{fig:dataset}.

In our setup, we have a single-antenna transmitter (TX) and a two-antenna receiver (RX). The TX can either act as a legitimate sender or a jammer. First, we generate random samples and save in  memory  both at the TX and RX. Then, the TX transmits using the saved data samples and the RX records the received RF samples from each of the two antennas to files. Due to the channel effects, the received samples are rotated by some unknown phase shift ($\phi_1$ and $\phi_2$ for the two antennas as shown in \Cref{fig:dataset}). These values are different for each of the receiving antennas. To determine these values, we chunk the received samples and cross-correlate with the data samples already saved in the RX’s memory from the first step. $\phi_1$ and $\phi_2$ are then computed from the angular values (argument) of the correlation outputs with the highest energy (peak).  With this approach, we determine the phase shift $\Delta_{\phi_S}$ (for the legitimate signal, or $\Delta_{\phi_J}$  for the jamming signal) by taking the difference of the two angles. When the communication channel is static, these phase shifts will experience very little variance. Such dataset would negatively affects the training and bias the resulting model to a small range of output values. We address this by devising a simple RF augmentation technique. For each chunk of RF samples, we rotate one antenna by a random value between $[-\pi,\pi]$ and adjust the phase shift calculated in the last step accordingly. This results in a more diverse dataset.

The above process of acquiring phase shifts is performed for both the sender and the jammer. After that, we get the data that represents collisions by adding the samples recorded for the sender and the jammer together, with the phase shifts $\Delta_{\phi_S}$ and $\Delta_{\phi_J}$ acquired from the previous process as the labels for phase shift estimation. We also collects the data representing noise with the TX turned off. Our final dataset comprises data and labels for all possible cases: When the channel is vacant, when it is occupied by one emitter, or when there are collisions between the sender and the jammer.

\subsection{Loss Function and Training}
\bfpara{Loss Function} During the training, our CNN aims to minimize a loss function which represents the errors of phase shift estimations and signal detections. For the phase shift estimations, we use a modified Mean Squared Error function:
\begin{equation}
\mathcal{L}_{\phi} = 1_{S_1}(\Delta_{\phi_1}-P_{S_1})^2 + 1_{S_2}(\Delta_{\phi_2}-P_{S_2})^2
\end{equation}
where $\Delta_{\phi_1},\Delta_{\phi_2}$ are the ground truth values and $P_{S_1}, P_{S_2}$ are the output estimations (shown in~\Cref{fig:cnn}). $1_{S_1}$ (or $1_{S_2}$) is 1 if $\Delta_{\phi_1}$ (or $\Delta_{\phi_2}$) associates with a signal, otherwise 0. For the signal detections, we use Binary Cross-Entropy loss function:
\begin{align}
\begin{split}
\mathcal{L}_{S} = - ((1_{S_1}\log(I_{S_1}) + 0_{S_1}\log(1-I_{S_1})) +\\
        (1_{S_2}\log(I_{S_2}) + 0_{S_2}\log(1-I_{S_2})))    
\end{split}
\end{align}
where $I_{S_1}, I_{S_2}$ are the outputs for signal detection associated respectively with $P_{S_1}, P_{S_2}$. Meanwhile, $0_{S_1}$ ($0_{S_2}$) is the complement of $1_{S_1}$ ($1_{S_2})$. The final loss function is the sum of the two loss components:
\begin{equation}\label{eq:combination_loss}
    \mathcal{L} = \mathcal{L}_{\phi}+\mathcal{L}_{S}
\end{equation}

\bfpara{Training} After a large number of iterations for validation, our CNN is finalized for the training. We use PyTorch library \cite{pytorch} to develop the CNN model. To improve the training convergence and eliminate the needs for regularization, we utilize Batch Normalization \cite{batchnorm} on the outputs of the convolutional layers. Furthermore, we minimize the possibility of overfitting by using a Learning Rate Decay technique \cite{rong2019} in which we lower the learning rate when the validation error does not improve over a period of time, e.g., a few training epochs. We emphasize that the phase shift estimations $P_{S_1}, P_{S_2}$ are made distinguishable during training by assigning them to learn the smaller and bigger phase shifts, respectively.

%% file: cancellation.tex
\section{Jamming cancellation}\label{sec:cancel}
In this section, we present the details of our jamming cancellation procedure that utilizes the phase shift estimations and signal detections from the CNN model. To perform the cancellation as discussed in \Cref{sec:approach}, we  devise the algorithms and formulas for estimating the amplitude ratio and for making the phase shift estimations more robust. The jamming cancellation algorithm is described in \Cref{alg:cancel}.

\begin{algorithm}[hbt!]
\caption{CNN-based Jamming Cancellation}\label{alg:cancel}
\SetAlgoLined
\SetKw{Or}{\hspace{0.3em}\itshape or \hspace{0.3em}}
\SetKw{And}{\hspace{0.3em}\itshape and \hspace{0.3em}}
\KwData{$P^T_{S_1}, P^T_{S_2}, 1^T_{S_1}, 1^T_{S_2}, \Delta^{cur}_{\phi_J}$, RF samples at time period $T$}
\KwResult{$I^{last}_J, E_{J_i}, E_{S_i}$ with $i \in \{1, 2\}$, RF samples}
\uIf{$1^T_{S_1} \bigoplus 1^T_{S_2} = 1$}{
    Decode the RF samples\;
    \uIf{decodable}{
        Measure the signal power $E^i_S$ with $i \in \{1, 2\}$\;
        $I^{last}_J \gets 0$\;
    }
    \Else{
        Measure the jamming power $E^i_J$ with $i \in \{1, 2\}$\;
        Calculate $\Delta_{\phi_J}$ using \Cref{eq:smoothing}\ and update $\Delta^{cur}_{\phi_J}$\;
        $I^{last}_J \gets 1$\;        
    }
}
\uElseIf{$1^T_{S_1} = 1 \And 1^T_{S_2} = 1$}{
    \uIf{$\left|P^T_{S_1}-\Delta^{cur}_{\phi_J}\right| < \left|P^T_{S_2}-\Delta^{cur}_{\phi_J}\right|$}{
        $\Delta^T_{\phi_J} \gets P^t_{S_1}$\; 
    }
    \Else{
        $\Delta^T_{\phi_J} \gets P^t_{S_2}$\; 
    }
    Calculate $\Delta_{\phi_J}$ using \Cref{eq:smoothing}\ and update $\Delta^{cur}_{\phi_J}$\;
    Calculate amplitude ratio using \Cref{alg:amp_est}\;
    Removing jamming signal by \Cref{eq:jam_cancel}\;
}
\Else{
    Skip the current period\;
}

\end{algorithm}

\subsection{Analyze CNN Outputs}\label{sec:analyze}
As described in the previous section, at time period $T$ the receiver collects a block of RF samples, which is subsequently inputted to the CNN model to output the estimations $P^T_{S_1}, P^T_{S_2}, I^T_{S_1}, I^T_{S_2}$. $P^T_{S_i}$ represents the phase shift estimation for the current signal, while $I^T_{S_i}$ classifies the type of the corresponding phase shift, i.e. real signal or  noise, with $i \in \{1, 2\}$. We remind the reader that we distinguish the estimations by the ordering $P^T_{S_1}<P^T_{S_2}$ as learned during the training. We define the \textit{signal detection} indicator $1^T_{S_i}$ for corresponding estimation $P^T_{S_i}$ using the output $E^T_{S_i}$:
\begin{equation}\label{eq:convert_output}
    1^T_{S_i} = \begin{cases}
        1 &\text{$E^T_{S_i} > 0.5$}\\
        0 &\text{otherwise}
\end{cases} \; \forall i \in \{1, 2\}
\end{equation}
$1^T_{S_i}$ being equal to $1$ or $0$ indicates that $S_i$ (whose phase shift is estimated by $P^T_{S_i}$) is a real signal or noise, respectively. We can therefore recognize the current state of the communication channel and subsequently acquire the correct phase shift for cancelling the jamming signal when collisions happen (as shown in \Cref{alg:cancel}).

\noindent \textit{\textbf{When only $1^T_{S_1}$ or $1^T_{S_2}$ is $\mathbf{1}$:}} This indicates that the channel is currently used by a single transmitter, which can be either the legitimate sender or the jammer. We identify the jammer by checking if the RF samples are decodable. In this case the capability of the jammer is limited to degrading the communications between the nodes by occupying the channel. If we identify the jammer's presence (samples are not decodable), the estimation $P^T_{S_i}$ where $1^T_{S_i}=1$ signifies the jamming signal.  In the case where the adversary transmits data that mimicks legitimate communications, a solution can consist of duplicating the receiver chain continuously tracking and decoding both inferred signals (at the expense of doubling the receiver cost). Smarter approaches are possible by tracking the phases of the transmitters of interest and canceling other ones.

\noindent \textit{\textbf{When both $1^T_{S_1}$ and $1^T_{S_2}$ are $\mathbf{1}$:}} In this case, we detect a collision indicating that the legitimate signal is being interfered with by a jammer. We identify the jamming phase shift $\Delta^T_{\phi_J}$ out of the two estimation outputs $P^T_{S_1}, P^T_{S_2}$ by calculating the distance to the existing jamming phase shift estimation $\Delta^{cur}_{\phi_J}$ and picking the one with the closest distance. This is based on the prior assumption of  slow-fading channel for our setup, where the phase shift varies  slowly over time. $\Delta^{cur}_J$ is calculated and updated by a smoothing process on the history data of jamming phase shift estimation, which will be described in \Cref{sec:smoothing}. We emphasize that our detection is effective even if the jammer transmits at low-power (i.e. the jamming signal is weaker than the legitimate signal). Therefore, this low-level of interference will also be canceled. 

Apart from estimating the jamming phase shift, we also need to estimate the amplitude ratio $A_J =\frac{\left|h_{J_1}\right|}{\left|h_{J_2}\right|}$ as discussed in \Cref{sec:approach}. We introduce the intuition behind our approach. We analyze the power variation of RF samples in the periods before and during the collision, presented in details in \Cref{sec:amp_est}. With the phase shift and amplitude ratio estimated, the receiver can solve \Cref{eq:jam_cancel} with $p_1=\frac{h_{J_1}}{h_{J_2}}=A_Je^{j\Delta_{\phi_J}}$ to null the jamming component in the received signal. The legitimate signal now has a new gain $p_2$ and can be used to decode the data. While $p_2$ is not necessary to estimate, as mentioned in \Cref{sec:approach}, we note that the phase shift separation $Sep_{\Delta_\phi}$ between signals $S$ and $J$ has a direct impact on the gain $p_2$ and subsequently the quality of the final signal. This impact will be measured and discussed in \Cref{sec:evaluation}.

\noindent \textit{\textbf{When both $1^T_{S_1}$ and $1^T_{S_2}$ are $\mathbf{0}$:}} This informs the receiver that neither communication nor jamming is happening in the channel and we can skip this period. The ability to identify this channel state helps the receiver to reduce the computational power and to avoid corrupting the phase shift estimation in the long run.

\begin{figure}
    \centering
    \includegraphics[width=\linewidth]{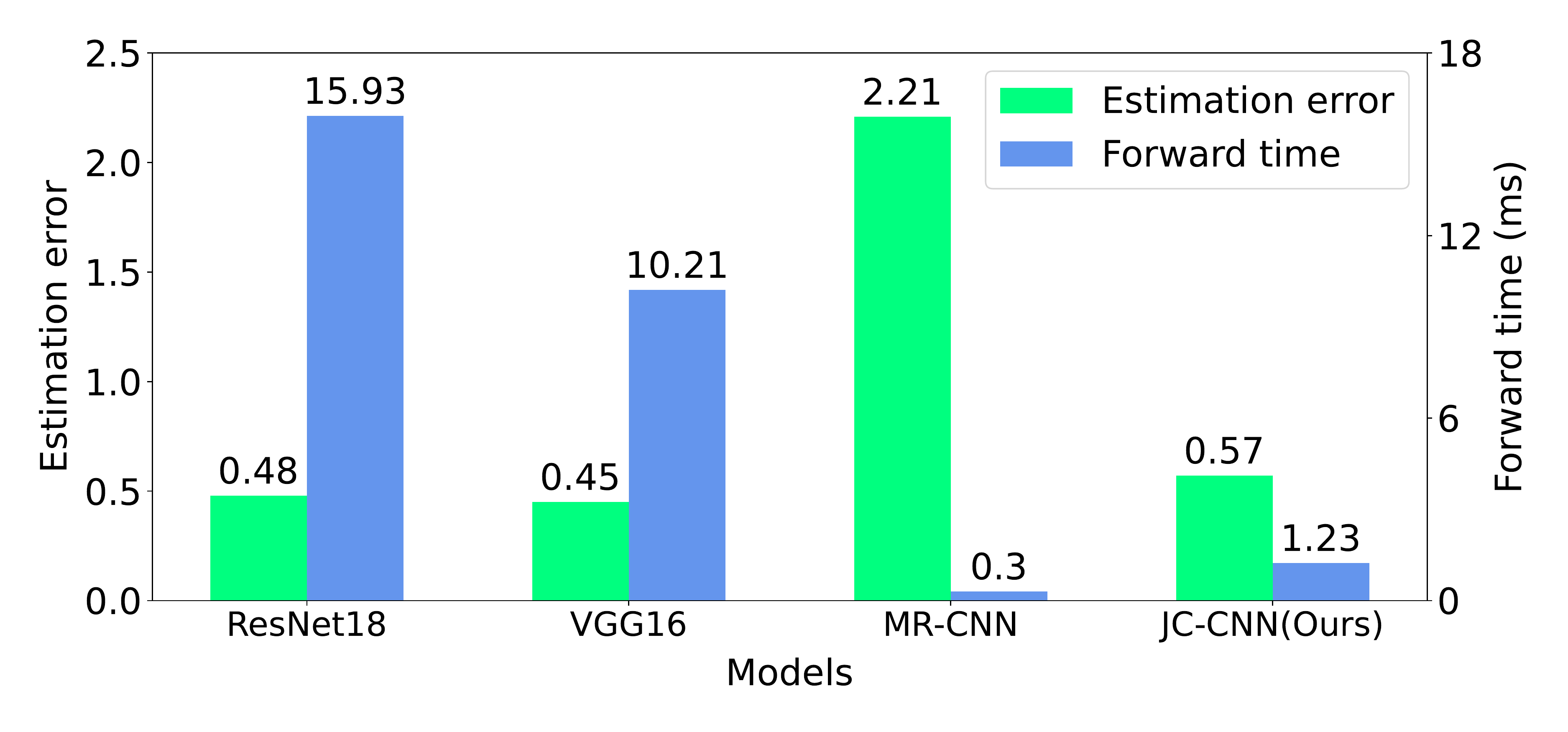}
    \caption{Comparison of CNN models for estimation error and network forward time.}
    \label{fig:cnn_comparison}
\end{figure}

\subsection{Amplitude Ratio Estimation}\label{sec:amp_est}

Our algorithm for estimating the amplitude ratio is described in \Cref{alg:amp_est}. It should be noted that as described in \Cref{sec:analyze}, \Cref{alg:amp_est} is only triggered when a collision is detected and the amplitude ratio needs to be estimated to perform the jamming cancellation. Our approach is inspired by the observation that the power of a receiving antenna during a collision comprises two independent, separable components for the legitimate signal $S$ and the jamming signal $J$. Suppose during time period $T$, the receiver collects $N$ digital RF samples from the analog input of antenna $i$, the received power $ E^T_i$ can be written as:
\begin{equation}\label{eq:cur_power}
    E^T_i=\frac{1}{N}\sum_{t=1}^N |h_{S_i}^t S^t + h_{J_i}^t J^t|^2
\end{equation}
Given that the channel is  slow-fading and not dependent on the instant time $t$, and because the sender's signal $S$ and the jammer's signal $J$ are uncorrelated, i.e. $\sum_{t=1}^N h^t_{S_i} S^t (h^{t}_{J_i} J^{t})^*=0$ and $\sum_{t=1}^N (h^t_{S_i} S^t)^* h^{t}_{J_i} J^{t}=0$, \Cref{eq:cur_power} becomes:
\begin{equation}\label{eq:sep_power}
    E^T_i=\frac{1}{N}(|h_{S_i}|^2 \sum_{t=1}^N|S^t|^2) + \frac{1}{N}(|h_{J_i}|^2 \sum_{t=1}^N|J^t|^2) = E_{S_i} + E_{J_i}  
\end{equation}
\ignore{where we assume that the power of signal $S$ and $J$ is considerable higher than the noise. }To estimate $A_J=\frac{\left|h_{J_1}\right|}{\left|h_{J_2}\right|}=\sqrt\frac{E_{J_1}}{E_{J_2}}$ (where $E_{J_i}=\frac{1}{N}(|h_{J_i}|^2 \sum_{t=1}^N|J^t|^2)$), we need to measure $E_{S_i}= \frac{1}{N}(\left|h_{S_i}\right|^2 \sum_{t=1}^N|S^t|^2) $ for two antennas $i \in \{1,2\}$. To do this, we first determine whether the legitimate sender or the jammer transmits first, right before the collision (using the detection capability described in \Cref{sec:analyze}), then calculate the power accordingly. Here, we assume the sender's power is stable during the transmission of a packet (slow fading channel), so the measurement of $E_{S_i}$ at the beginning of the collision can be used until the end of that collision. Our algorithm looks at period $T-1$ right before the collision, and identifies the transmitter in that period with parameter $I^{last}_J$ (which is updated in \Cref{alg:cancel} and utilized in \Cref{alg:amp_est}). If the sender appears in period $T-1$ (where $I^{last}_J=0$), we can measure $E_{S_i}$ and calculate $A_J$. Otherwise, we know that the jammer appears in period $T-1$, and we can measure $E_{J_i}$ and update $E_{S_i}$ with the current power $E_i$. It should be noted that in the latter case, the new $E_{S_i}$ is used until the end of the collision and should not be updated again if we detect a collision in the previous period.

\begin{algorithm}[hbt!]
\caption{Amplitude Ratio Estimation}\label{alg:amp_est}
\SetAlgoLined
\SetKwComment{Comment}{/* }{ */}
\SetKw{Or}{\hspace{0.3em}\itshape or \hspace{0.3em}}
\SetKw{And}{\hspace{0.3em}\itshape and \hspace{0.3em}}
\KwData{$E_{J_i}, E_{S_i}, 1^{T-1}_{S_i}$ with $i \in \{1, 2\}$, $I^{last}_J$, RF samples at $T$}
\KwResult{$A_J$}
Measure the current power $E^T_i$ with $i \in \{1, 2\}$\;
\If{$I^{last}_J = 1 \And 1^{T-1}_{S_1} \bigoplus 1^{T-1}_{S_2} = 1$}
{
    $E_{S_1} \gets E^T_1 - E_{J_1}$\;
    $E_{S_2} \gets E^T_2 - E_{J_2}$\;
}
$A_J \gets \sqrt{\frac{E^T_1 - E_{S_1}}{ E^T_2 - E_{S_2}}}$ 
\end{algorithm}


\subsection{Estimation Smoothing}\label{sec:smoothing}
Unlike the estimations $I^T_{S_1}, I^T_{S_2}$ which are discretized to the values of $0$ and $1$ for the signal detections, the real-valued estimations $P^T_{S_1}, P^T_{S_2}$ are used directly to solve the jamming cancellation equation. This makes the cancellation process susceptible to the neural network's estimation variations and outliers \cite{bishop_prml}. We improve the robustness of the phase shift estimation and the subsequent jamming cancellation process by stabilizing the estimations with the exponential smoothing function:
\begin{equation}\label{eq:smoothing}
    \Delta_{\phi_J} = \Delta^T_{\phi_J} \lambda + \Delta^{cur}_{\phi_J} (1-\lambda)
\end{equation}
where $\lambda$ controls the smoothness of the output. We note that after performing the cancellation for the current period, $\Delta^{cur}_{\phi_J}$ is updated to the current value of $\Delta_{\phi_J}$. With this, the phase shift estimation $\Delta_{\phi_J}$ is more stable which makes the jamming cancellation process more robust. We will show how the smoothing algorithm and $\lambda$ parameter impact the performance of the cancellation through experimental evaluations in \Cref{sec:evaluation}.







%% file: evaluation.tex
\section{Evaluation}\label{sec:evaluation}
In this section, we evaluate our CNN-based jamming cancellation approach in multiple settings of modulations and wireless channels. Using the techniques mentioned in \Cref{sec:data_collection}, we collect RF samples transmitted by a sender and a jammer through coaxial cables to a receiver, and build a dataset containing $7,648,260$ real-valued data tensors of size $2 \times 128 \times 2$ reflecting $I/Q$ values of 128 RF samples collected by two antennas. Each data tensor is labeled by the phase shifts of signals embedded in the data (if the data is collected from the sender, or the jammer, or the combination of the two) and the signal indicators ($0$ for noise and $1$ for real signal) for their respective phases. The transmitters and receiver are implemented using GNURadio \cite{gnuradio} running on ETTUS USRP B210 software-defined radios. The radios use differential BPSK, QPSK, 8-PSK and 16-QAM modulation schemes to transmit and receive at a center frequency of $795$ MHz. We split the dataset into three parts used for training, validation, and testing with the ratio $0.64: 0.16 : 0.2$, respectively. Our trained CNN model is used for both over-the-cables evaluation (\Cref{sec:over_cable}) and over-the-air evaluation (\Cref{sec:over_air}). We emphasize that while the CNN model is trained only using the data collected through cables, our jamming cancellation approach still performs well in over-the-air indoor environment with the presence of multi-path and other channel effects not experienced in the cabled environments.  

\begin{figure}
    \centering
    \subcaptionbox{Bit Error Rate. \label{fig:smoothing_ber}}{
        \includegraphics[width=\linewidth]{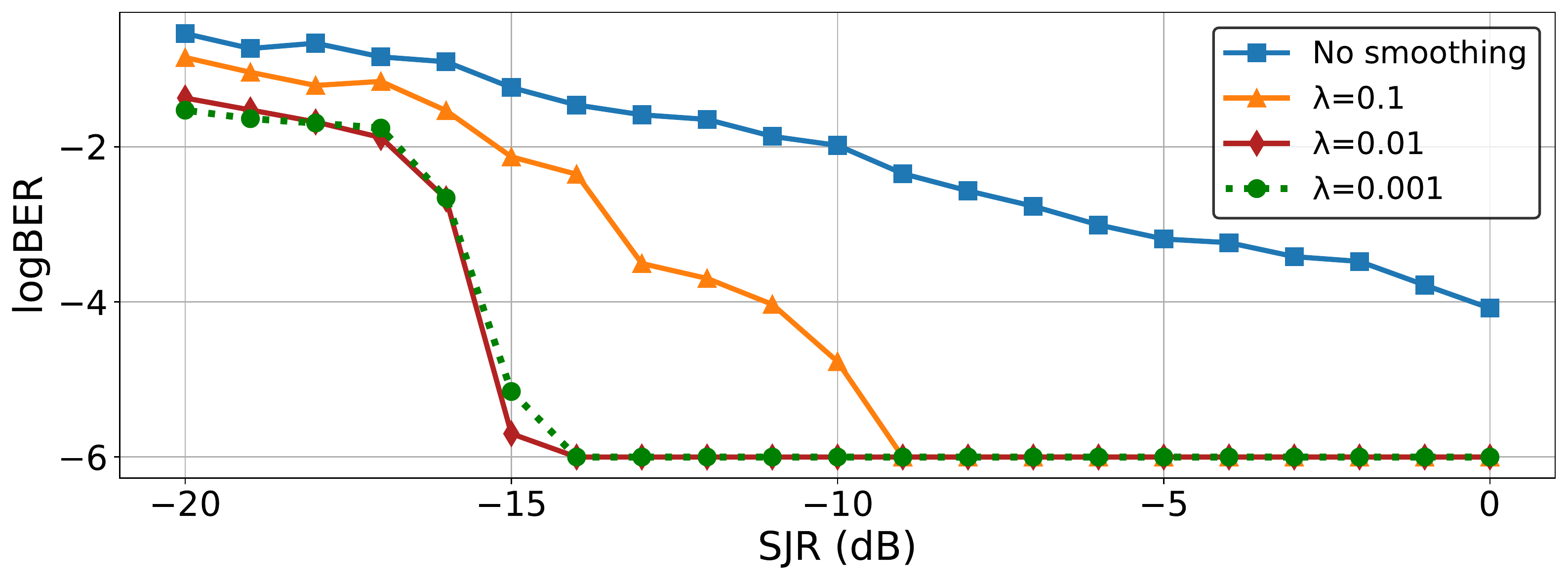}
    }
    \subcaptionbox{Energy variation.\label{fig:smoothing_energy}}{
        \includegraphics[width=\linewidth]{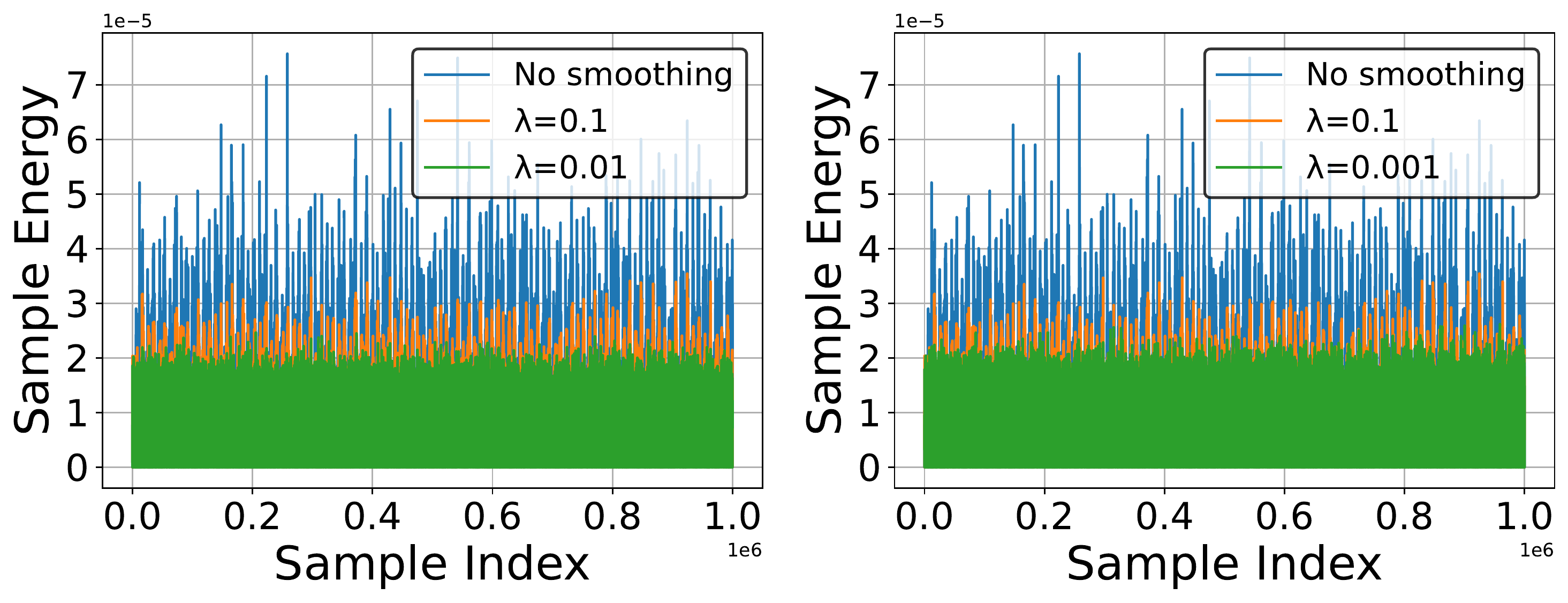}
    }
    \caption{Estimation smoothing reduces energy fluctuation from cancellation and improves Bit Error Rate. Decreasing $\bm{\lambda}$ from $\bm{0.01}$ to $\bm{0.001}$ does not further improve the energy variation (Plot (b): $\bm{\lambda=0.01}$ (left) has similar variation to $\bm{\lambda=0.001}$ (right)). Similarly, for the Bit Error Rate (Plot (a)).}
    \label{fig:smoothing}
\end{figure}

\begin{figure*}
    \centering
    \subcaptionbox{BPSK}{
        \includegraphics[width=.45\linewidth]{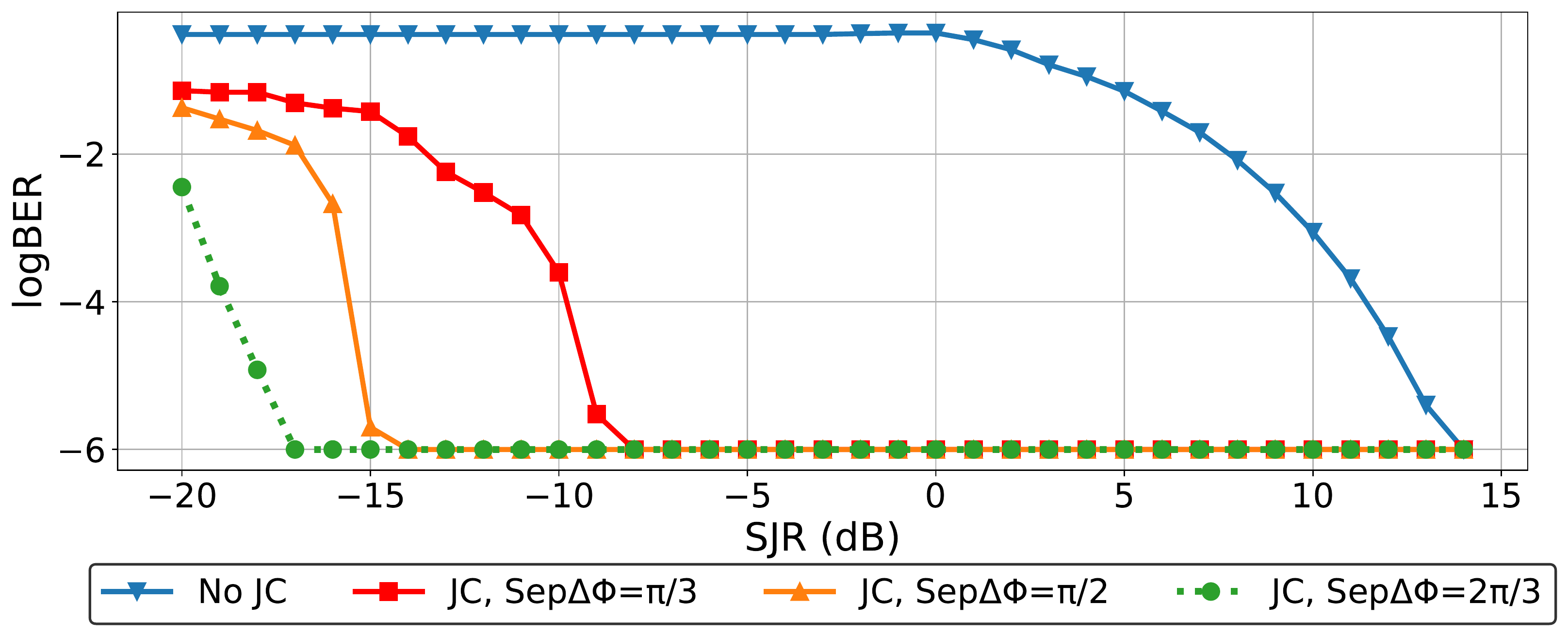}
    }
    \subcaptionbox{QPSK}{
        \includegraphics[width=.45\linewidth]{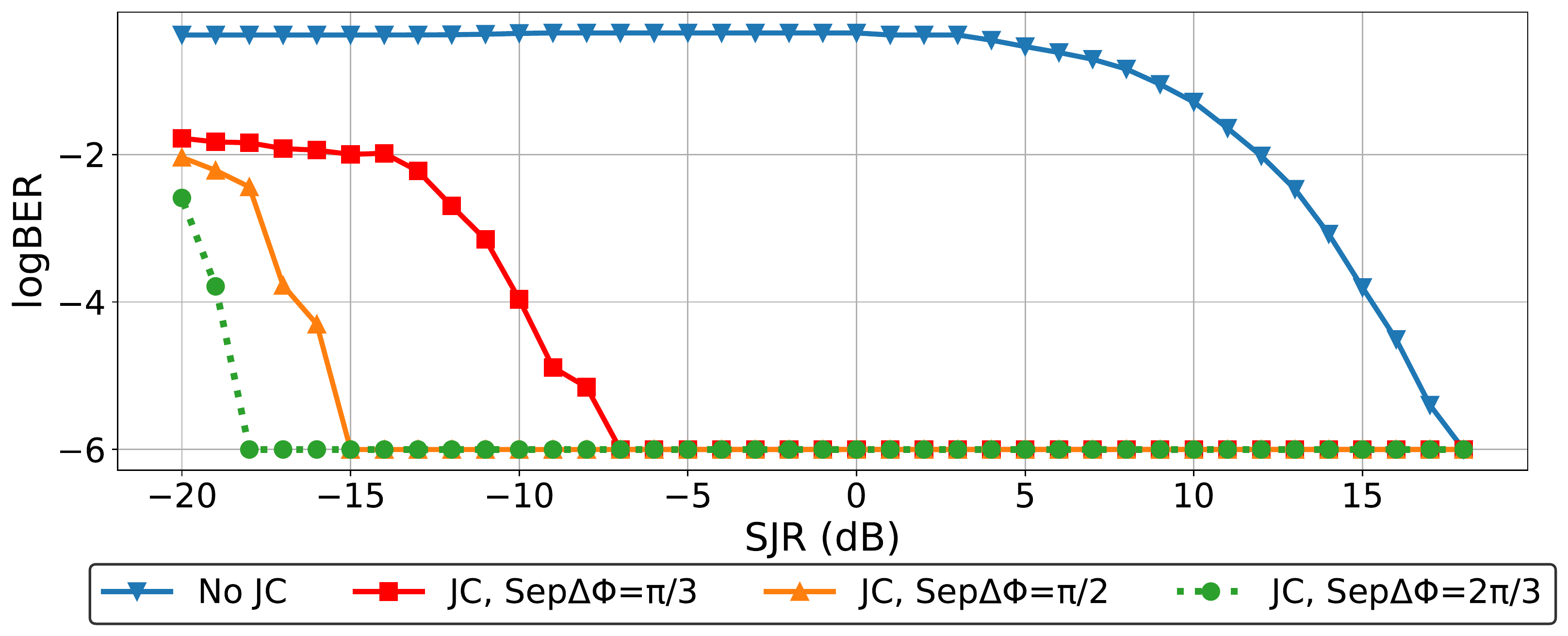}
    }
    \subcaptionbox{8-PSK}{
        \includegraphics[width=.45\linewidth]{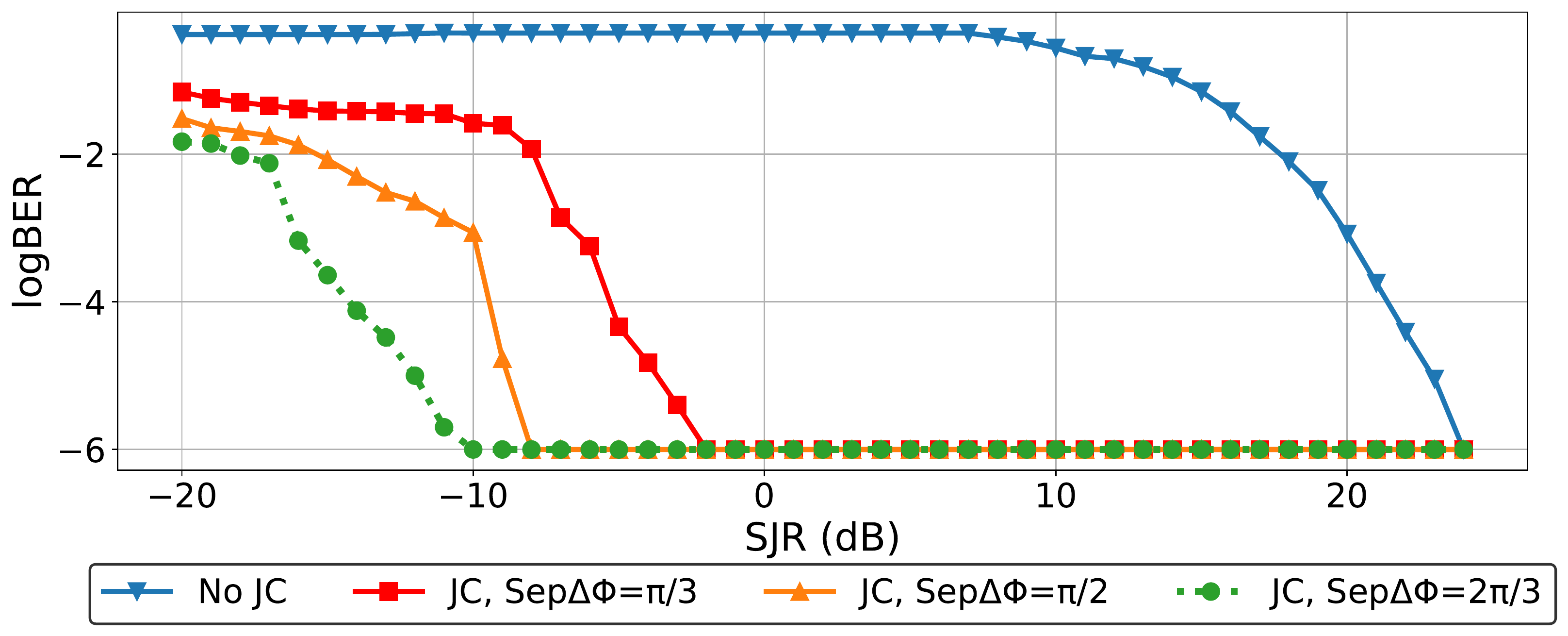}
    }
    \subcaptionbox{16-QAM}{
        \includegraphics[width=.45\linewidth]{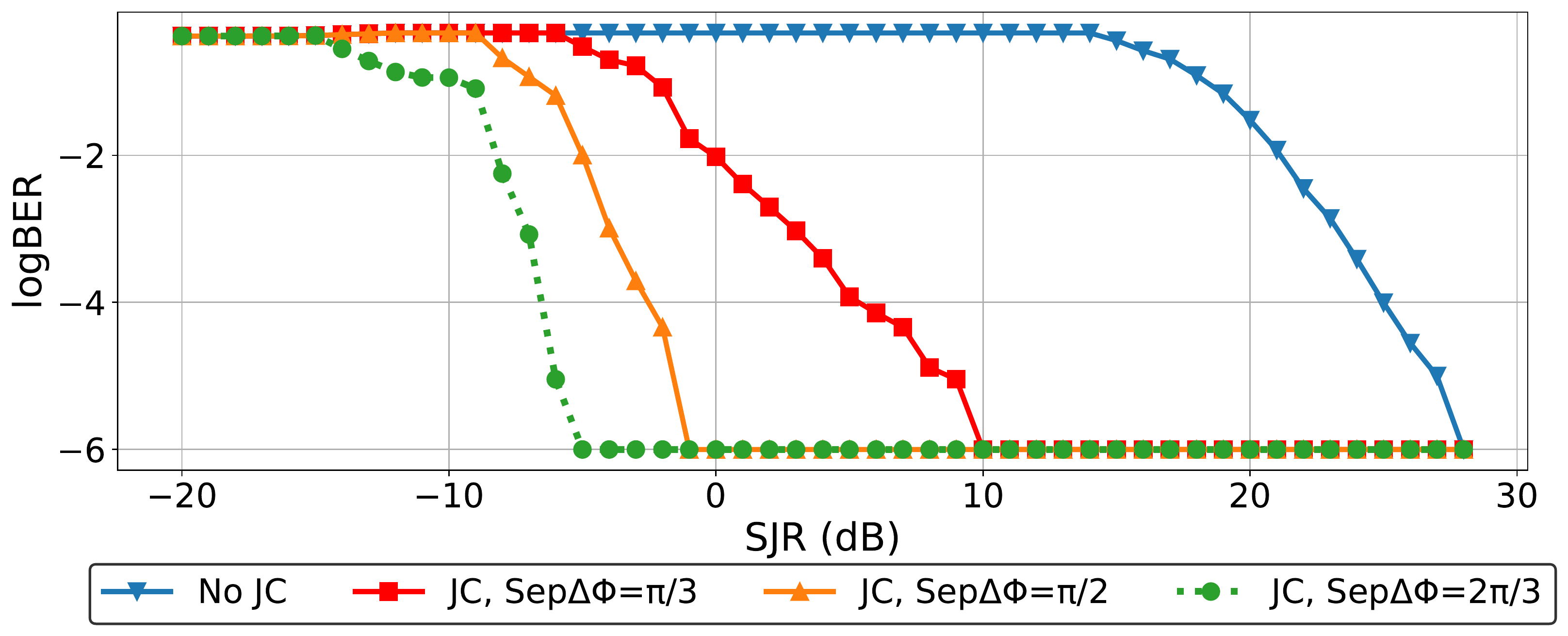}
    }

    \caption{Bit Error Rate evaluation for over-the-cables experiments.}
    \label{fig:ota_eval}
\end{figure*}

\subsection{Neural Network Comparison}
Beyond evaluating the performance of our novel approach, we are also interested in understanding the performance of the specific CNN architecture we proposed. Towards this goal, we compare it with  existing CNN models from the literature, in terms of estimation error and network forward time (i.e., the elapsed time from when the network receives data to when it outputs estimations). In our setup, we train our CNN model (called JC-CNN), VGG16~\cite{VGG-2014}, ResNet~\cite{resnet16}, and the modulation recognition CNN (MR-CNN)~\cite{oshea16} and evaluate using the training and test set of our jamming detection and cancellation dataset, respectively. The models are implemented, trained, and tested using the Pytorch library~\cite{pytorch} and CUDA~\cite{cuda} Version 10.2 running on a NVIDIA GeForce GTX 1080 GPU. We use Adam Optimizer \cite{kingma2017adam} and~\code{ReduceLROnPlateau} learning rate scheduler~\cite{yogi} with initial learning rate $lr=0.005$ for the training of all models. The metric for estimation error is the combination loss denoted by \Cref{eq:combination_loss}. To benchmark the forward time, we use Pytorch's \colorbox{codegray}{\code{torch.cuda.synchronize}} wrapping around neural network's forward propagation function to synchronize CUDA operations for accurate timing measurement. The final test loss and forward time are achieved by averaging over $20,000$ iterated measurements, and illustrated in \Cref{fig:cnn_comparison}. It is clear that our JC-CNN model outperforms MR-CNN in the correctness of the estimations, with the test error of $0.57$ compared to $2.21$ of MR-CNN (over $3.8$ times lower), while is slightly less accurate than VGG-16 and ResNet18 with the test losses of $0.45$ and $0.48$, respectively. However, our model is $12.95$ times faster than ResNet18 and $8.3$ times faster than VGG16. Therefore, the JC-CNN model has a better optimization of speed and accuracy for this task and is more suitable to deploy for real-time and embedded applications.

\subsection{Over-The-Cables Evaluation}\label{sec:over_cable}
First, we evaluate the efficiency of the jamming cancellation approach in a relatively idealistic environment where RF signals propagate through coaxial cables, thus multi-path and other fading effects are absent. Our setup comprises a sender, a receiver and a jammer, where the sender and the jammer transmit modulated signals using differential BPSK, QPSK, 8-PSK and 16-QAM. The efficiency of jamming cancellation is measured by the Bit Error Rate metric, which we calculate by comparing and counting the error bits between the sent and the received signals (signals are recorded at the nodes and transferred to the host computer for calculation). 

In \Cref{sec:approach} we show that the phase shift separation $Sep_{\Delta_\phi}=\left|\Delta_{\phi_S}-\Delta_{\phi_J}\right|$ being very small can cause negative effects to the legitimate signal even when the jamming signal is completely removed. In our experiment, the transmitters are connected to the sender by identical coaxial cables, in which $Sep_{\Delta_\phi} \approx 0$. To address this, we introduce an artificial channel effect by shifting the phases of the antennas. Depending on the shifting, $Sep_{\Delta_\phi}$ will receive a different value. We discuss the impact of  $Sep_{\Delta_\phi}$ on the efficiency of the jamming cancellation in the evaluation below.

\begin{figure*}\label{fig:cab_eval}
    \centering
    \subcaptionbox{$1^{st}$ configuration}{
        \includegraphics[width=.45\linewidth]{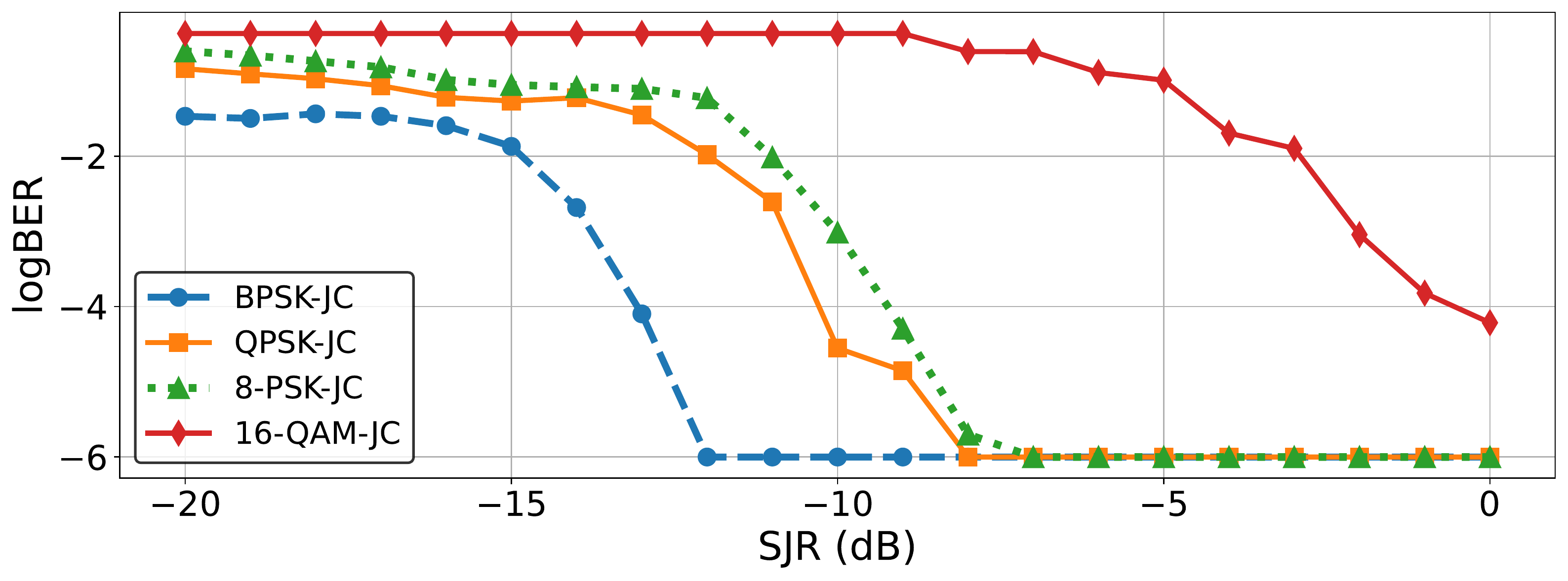}
    }
    \subcaptionbox{$2^{nd}$ configuration}{
        \includegraphics[width=.45\linewidth]{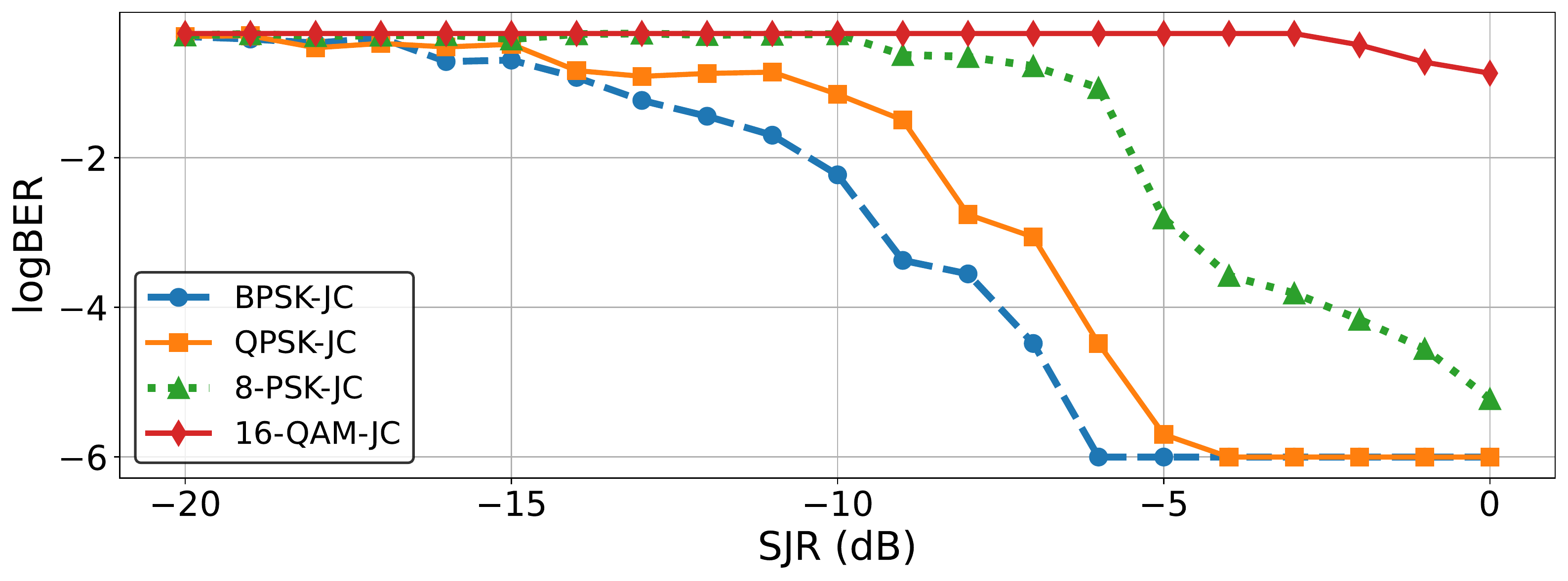}
    }
    \subcaptionbox{$3^{rd}$ configuration}{
        \includegraphics[width=.45\linewidth]{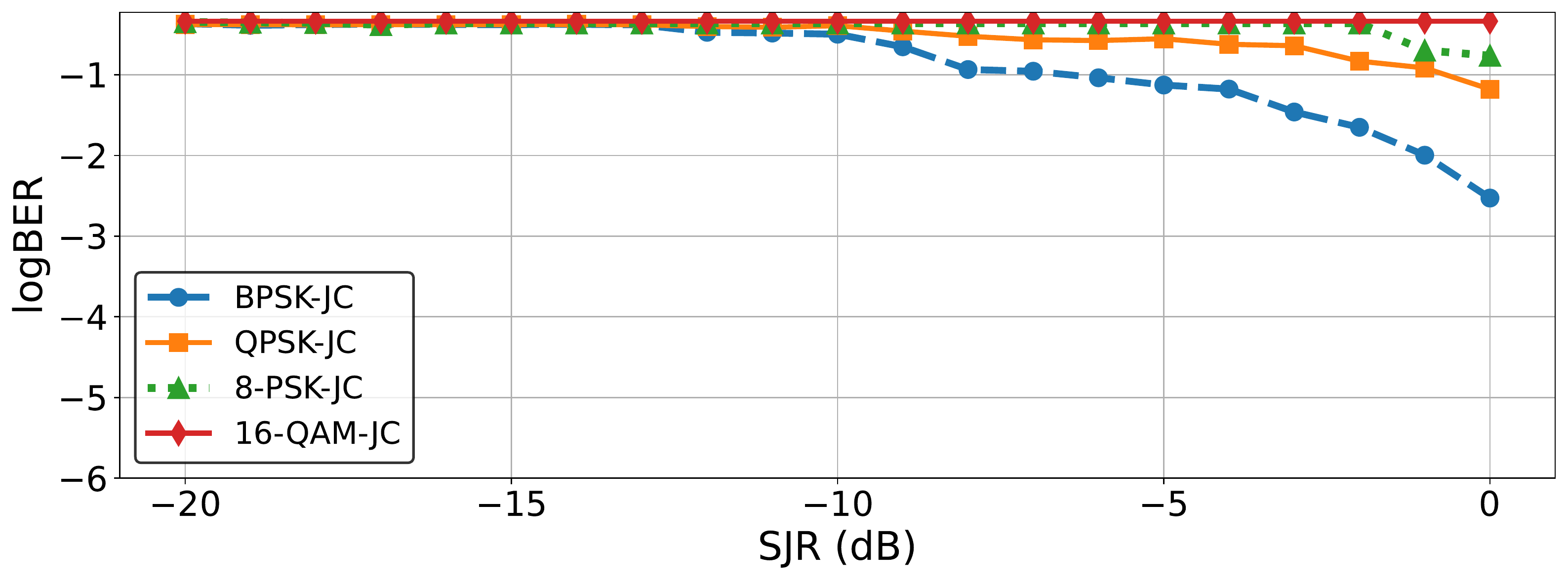}
    }
    \subcaptionbox{$4^{th}$ configuration}{
        \includegraphics[width=.45\linewidth]{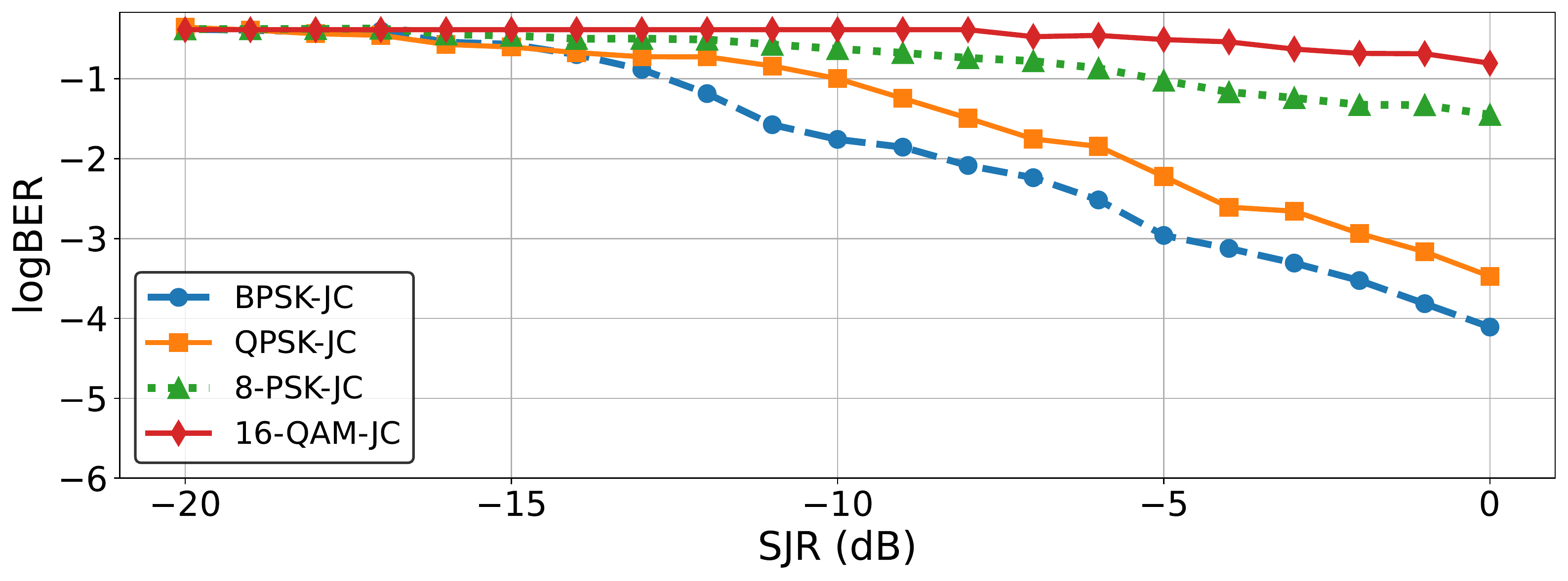}
    }

    \caption{Bit Error Rate evaluation for over-the-air experiments. When the legitimate signal and the jamming signal are phase-aligned (i.e. $\bm{Sep_{\Delta_\phi} \approx 0}$), removing the jamming signal from the received signal also heavily degrades the legitimate signal and impacts the Bit Error Rate.}
    \label{fig:ota_eval}
\end{figure*}

\bfpara{Impact of Phase Separation} \Cref{fig:cab_eval} shows the Bit Error Rate (BER) evaluation considering four cases: No jamming cancellation ($JC$) is applied, and jamming cancellation is applied with three values of $Sep_{\Delta_\phi}: \frac{\pi}{3}, \frac{\pi}{2},$ and $\frac{2\pi}{3}$. In this evaluation, the jamming cancellation algorithm uses the estimation smoothing with parameter $\lambda=0.01$. First, it is clear to see that our cancellation approach can achieve very high jamming resistance: It allows the receiver to operate at BER of $10^{-6}$ with the Signal-to-Jamming Ratio of $-18$dB (i.e., the jammer is 63 times more powerful than the legitimate signal) for QPSK with $Sep_{\Delta_\phi}=\frac{2\pi}{3}$. Interestingly, \textit{when compared with the case of no cancellation, our approach achieves over $30$dB gain for all modulations when operating at a BER of under $10^{-4}$} (e.g. $31$ dB for BPSK and $34$ dB for QPSK with $Sep_{\Delta_\phi}=\frac{2\pi}{3}$). In addition, we also see that the jamming cancellation performs better as $Sep_{\Delta_\phi}$ gets bigger. For instance, the jamming resistance when operating at a BER of $10^{-6}$ with BPSK modulation drops by $3$dB when $Sep_{\Delta_\phi}$ decreases from $\frac{2\pi}{3}$ to $\frac{\pi}{2}$, and by $9$dB when it decreases to $\frac{\pi}{3}$. It is easy to see the same trend for the other modulations. This limitation is intrinsic to multi-antenna jamming cancelation as the receiver cannot resolve two transmitter that are aligned with it. Furthermore, we also see that the efficiency of our jamming cancellation when using 8-PSK and 16-QAM is lower compared to BPSK and QPSK, which is expected because they have smaller distance between the constellation points and thus are more prone to bit errors~\cite{proakis2008digital}. 

\bfpara{Impact of Estimation Smoothing}
We investigate the impact of phase shift estimation smoothing on the jamming cancellation with the Bit Error Rate evaluation shown in \Cref{fig:smoothing_ber}. In this case, we use BPSK signals and $Sep_{\Delta_\phi}=\frac{\pi}{2}$. It is clear to see that the smoothing significantly improves the jamming resistance: We achieve $11$ and $15$ dB gain with $\lambda=0.1$ and $0.01$ when operating at BER below $10^{-4}$, respectively. This effect can also be seen in \Cref{fig:smoothing_energy}, where the estimation smoothing helps stabilizing the energy of the samples and reduce both the degree and the frequency of energy variation, resulting in better BER. Finally, setting $\lambda$ to $0.01$ makes the energy more stable and yields better jamming resistance ($-14$ dB of SJR compared to $-9$dB for $\lambda=0.1$ when operating at BER=$10^{-6}$, while decreasing $\lambda$ to $0.001$ does not improve the performance further. Therefore, we selected $\lambda=0.01$ for all  later evaluations.

\begin{figure}
    \centering
    \includegraphics[width=.57\linewidth]{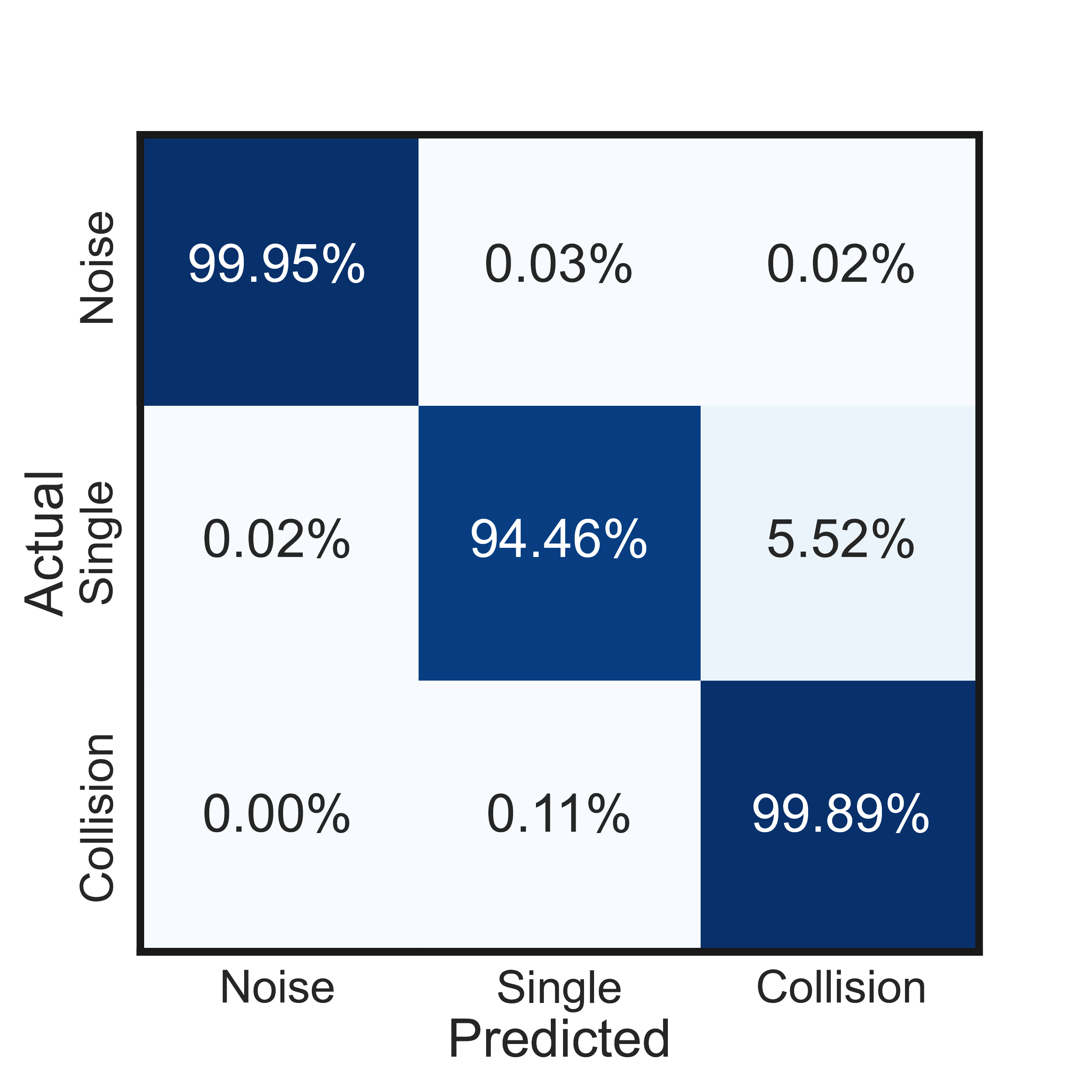}
    \caption{Confusion matrix of channel state classification.}
    \label{fig:confusion_matrix}
\end{figure}

\subsection{Over-The-Air Evaluation}\label{sec:over_air}
\bfpara{Jammer Detection} 
We conduct over-the-air experiments to assess the ability of our CNN model to detect a jammer detection, in an environment different from the one used for training, i.e., model trained on data recorded through cables is evaluated for over-the-air without re-training. Similar to over-the-cables experiments, our setup consists of a sender, a receiver, and a jammer. We use differential BPSK, QPSK, 8-PSK and 16-QAM modulations for the transmissions. The testbed is positioned in an indoor environment, where there are common RF-blocking and reflecting objects such as a PC, monitor, walls, and desks. To evaluate the detection capability, we focus on the classification of three channel states: (1) When there is no transmission and the channel is clear, (2) when there is a single transmission, and (3) when there are two transmissions (from the sender and the jammer) causing collisions. We note that in the second case, the transmitter being the sender or the jammer is decided by the decoding check in \Cref{alg:cancel}. \Cref{tab:state_clf} and \Cref{fig:confusion_matrix} depict the classification results, in which the CNN classifies three states: Noise (no transmission), Single (one transmission) and Collision (two transmissions). Our CNN model achieves 97.44\% accuracy, where it can reach over $99\%$ accuracy for the Noise and Collision states while the accuracy of Single state prediction is only about $5\%$ lower. We also get high scores for other metrics, over 0.98 for both Precision, Recall and F1-Score. The results justify the capability of our CNN model to identify the current channel state, and more importantly, the presence of a jammer (by recognizing collisions with $99.89\%$ accuracy) in the realistic environment without the needs to retrain the model trained in the idealistic environment (i.e. coaxial cables).

\begin{table}[t]
\centering
\begin{tabular}{|c|c|c|c|} 
 \hline
 Accuracy & Precision & Recall & F1-Score \\ 
 \hline
 97.44\% & 0.9818 & 0.981 & 0.9809 \\ 
 \hline
\end{tabular}
\caption{Channel classification results with different metrics.}
\label{tab:state_clf}
\end{table}

\bfpara{Jamming Resistance} To demonstrate the jamming resistance capability of our approach in over-the-air environments, we devise four setup configurations for our experiments. In each configuration, we randomly position the jammer, while the locations of the sender and the receiver remain fixed. To evaluate, we measure the Bit Error Rate (BER) of the received data for every case. \ignore{ We use the same BER calculation method as for over-the-cables experiments.} As discussed in the previous sections, different configurations will introduce different phase shift separations $Sep_{\Delta_{\phi}}$ that impact the quality of the signal resulting from the jamming cancellation. To measure $Sep_{\Delta_{\phi}}$ for each configuration, we use the cross-correlation technique presented in \Cref{sec:data_collection}. The BER evaluation results in \Cref{fig:ota_eval} show that the receiver equipped with the jamming cancellation ($JC$) achieves the best jamming resistance for both BPSK, QPSK, 8-PSK and 16-QAM modulations with the $1^{st}$ configuration whose phase shift separation is biggest (measured in \Cref{tab:config}). More specifically, we can operate with BPSK at BER of $10^{-6}$ under a Signal-to-Jamming Ratio (SJR) of $-12$ dB in the $1^{st}$ configuration, and under a SJR of -6 dB in the $2^{nd}$ configuration (which has the second-largest $Sep_{\Delta_{\phi}}$), while we only achieve the BER of down to $10^{-3}$ and $10^{-4}$ under a SJR of $0$ dB for the $3^{rd}$ and $4^{th}$ configurations, respectively. In the $3^{rd}$ configuration when the sender's signal and the jammer's signal are nearly phase-aligned ($Sep_{\Delta_{\phi}}$ has the lowest value of $0.34$), the resulting signal is especially weak and causes a high BER, i.e., over $10^{-2}$ for all modulations at SJR = $0$ except BPSK. This impact of $Sep_{\Delta_{\phi}}$ justifies the theory in \Cref{sec:approach} and our findings in over-the-cables experiments. Furthermore, we can also see that the efficiency of the jamming cancellation decreases as the modulation order increases. More specifically, compared to the case of BPSK modulation, using QPSK drops the jamming resistance by $2-3$dB, while the drop is $4-5$dB for 8-PSK and over $10$dB for 16-QAM. This is expected because higher-order modulations have a smaller distance between the constellation points that cause higher probability of bit errors \cite{proakis2008digital}. However, we note that the BER when not using our Jamming Cancellation is $\approx 0.5$ (theoretical maximum) for all the transmission settings and configurations shown in \Cref{fig:ota_eval}. Therefore, the results indicate the efficiency of our jamming cancellation approach, as well as the \textit{universality} for anti-jamming multiple modulation and in different environments not present during the training of the CNN model.


\begin{table}[t]
\centering
\begin{tabular}{|c|c|c|c|c|} 
 \hline
 Configuration & $1^{st}$ & $2^{nd}$ & $3^{rd}$ & $4^{th}$ \\ 
 \hline
 $Sep_{\Delta_\phi}$\ & 2.22 rad & 1.44 rad & 0.34 rad & 0.56 rad \\ 
 \hline
\end{tabular}
\caption{$Sep_{\Delta_\phi}$ measurements of over-the-air configurations.}
\label{tab:config}
\end{table}

%% file: related.tex
\section{Discussion and Related work}\label{sec:related}
The results show that our approach can perform well under the impact of multi-path, albeit \Cref{eq:recv_sig_jam_multi} only shows a single channel gain for either the jammer or the sender in the received signal. Nonetheless, the results do not contradict the cancellation theory. As explained in \cite{atj_mechanical_ws13}, the sum of the channel gains of all the paths from the sender (or the jammer) to the receiver can be viewed as a new channel gain of the line-of-sight path between the receiver and the sender/jammer being put in a different location. Therefore,~\Cref{eq:recv_sig_jam_multi} is also applicable in this case, and the jamming cancellation is still effective. 

It is demonstrated that the performance of multi-antenna jamming cancellation systems degrades when the emitters are phase-aligned. The next challenge is how to ensure a desirable phase shift separation to distinguish between two emitters and cancel the unwanted signal. To address this, it is intuitive to use and optimize a larger antenna array to exploit the diversity of multi-antenna and enhance the robustness of cancellation. On that account, exploring optimized settings of antenna array for robust jamming cancellation would be an interesting direction for our future work.

Traditional anti-jamming at the physical layer has been relying on spread spectrum techniques, which require the coordinating nodes to pre-share a secret key. Recent research efforts have addressed that limitation for FHSS~\cite{atj_control_channel_wisec09, atj_fhss_sp_08, atj_fhss_mobihoc_09, atj_fhss_tifs12}, or DSSS~\cite{atj_dsss_usenix09, atj_dsss_infocom10}, or both~\cite{atj_mobihoc09}. Nonetheless, these approaches are designed with the specific goal to remove the pre-shared secret and not to counter powerful jammers, i.e., a few orders stronger than the sending node.

There has been efforts on multi-antenna anti-jamming designed specifically for MIMO system~\cite{atj_mimo_infocomm14, atj_mimo_cns}. Compared to those approaches, our work is unique in the sense that we do not need training sequences or pilots for channel estimation. In a recent work~\cite{atj_mechanical_ws13}, the authors develop a hybrid anti-jamming system that utilizes mechanical antenna steering and multi-antenna software cancellation. While this approach achieves high jamming resiliency, the efficiency of the cancellation relies heavily on the after-effect of mechanical steering to the received jamming signal, while our CNN-based cancellation can operate even when the interference remains powerful.

In addition, researchers have spent significant efforts mitigating jammers at higher layers such as MAC~\cite{awerbuch2008jamming,richa2010jamming}, network layer~\cite{dong2009practical}, cross-layer~\cite{chiang2010cross} or timing channel between datalink and network layers \cite{xu2008anti}. Nonetheless, the need for an efficient, resilient anti-jamming technique for physical layer security is still very important because of the fact that high-power jammers are increasingly easy to build nowadays. 

Advances in Machine Learning and Deep Learning have been utilized in various domains including anti-jamming. In~\cite{atj_rl_icassp17}, Deep Reinforcement Learning is used to derive an optimal frequency hopping strategy to evade jammers. In~\cite{atj_gan}, the authors investigate Generative Adversarial Network for both jamming strategies and defense. Meanwhile, the anti-jamming capability of Convolutional Neural Networks - a vital Deep Learning building block, remains unexplored. Convolutional Neural Networks were successful in various tasks of wireless communications, such as modulation recognition~\cite{oshea16}, and RF emissions detection and classification~\cite{nguyen2021wideband}. To the best of our knowledge, our anti-jamming approach is the first work in the literature that utilizes CNNs to detect and cancel jammers efficiently. We achieve over $99\%$ accuracy for detecting jammers and can enhance a RF receiver to achieve a Bit Error Rate as low as $10^{-6}$ while facing an adversary at 18dB higher power than the legitimate signal. Our approach is agnostic to the communication link, i.e., does not require modifications to the link modulation, thus can be used as an \textit{universal} anti-jamming and interference-cancelling module for existing technologies and systems.

\ignore{The performance jamming cancellation algorithm relies on the CNN in two aspects: the accuracy of the signal detections and the precision of the phase shift estimations. Since they are not perfect, we need a means to make the system resilient to the estimation errors. We leave this unsolved question to the future work.}

\ignore{Both over-the-cables and over-the-air show that the jamming cancellation capability is directly affected by $Sep_{\Delta_\phi}$. While it is the limitation of the two-antenna system, we reckon that a more robust system can be achieved with a larger antenna array which has more nulling capability (with narrower beam pattern)... }

\ignore{Through this work, we show that the CNN model trained on data collected in a nearly perfect environment (through coaxial cables) can also provide good performance for realistic over-the-air environment. This strengthens our motivation of using CNN to capture and utilize robust channel characteristics that generalize for real-life wireless environment. Nonetheless, this does not decline the possibility of a better approach that can achieve more desirable results for a more diverse set of wireless channels. Diving deep into the architecture and optimizing for generalization would be an interesting direction for our future work. }

\balance